\documentclass[aps,prd,groupedaddress,showpacs]{revtex4}

\usepackage{graphicx}
\usepackage{dcolumn}
\usepackage{bm}
\usepackage{amsmath}
\usepackage{epstopdf}
\usepackage{amsfonts}
\usepackage{amssymb}
\usepackage{hyperref}
\usepackage{xcolor}
\usepackage{caption}
\usepackage{subcaption}
\usepackage{float}%Ubicacion_figuras

\usepackage{natbib}

\providecommand{\U}[1]{\protect\rule{.1in}{.1in}}

\newcommand{\baa}{\begin{align}}
\newcommand{\eaa}{\end{align}}

\newcommand{\be}{\begin{equation}}
\newcommand{\ee}{\end{equation}}
\newcommand{\bea}{\begin{eqnarray}}
\newcommand{\eea}{\end{eqnarray}}

\begin{document}

\title{Discriminating interacting dark energy models using Statefinder diagnostic}

\author{Raul Carrasco}
\affiliation{Instituto de F\'isica, Pontificia Universidad Cat\'olica de Valpara\'iso, Avenida Brasil 2950, Casilla 4059, Valpara\'iso, Chile.}
\email{raul.carrasco.v@mail.pucv.cl}

\author{\'Angel Rinc\'on}
\affiliation{Departamento de Física Aplicada, Universidad de Alicante, Campus de San Vicente del Raspeig, E-03690 Alicante, Spain}
\email{angel.rincon@ua.es}

\author{Joel Saavedra}
\affiliation{Instituto de F\'isica, Pontificia Universidad Cat\'olica de Valpara\'iso, Avenida Brasil 2950, Casilla 4059, Valpara\'iso, Chile.}
\email{joel.saavedra@pucv.cl}

\author{Nelson Videla}
\affiliation{Instituto de F\'isica, Pontificia Universidad Cat\'olica de Valpara\'iso, Avenida Brasil 2950, Casilla 4059, Valpara\'iso, Chile.}
\email{nelson.videla@pucv.cl}

\date{\today}

\begin{abstract}
In the present work, we perform a comparative study of different
interacting dark energy (DE) models using the 
Statefinder diagnostics. In particular, 17 different forms of the  energy transfer rate $Q$ between DE and dark matter (DM) were focused on, belonging to the following categories:
i) linear models in energy densities of DE and DM, 
ii) non-linear models, 
iii) models with a change of direction of energy transfer between DE and DM, 
iv) models involving derivatives of the energy densities, 
v) parametrized interactions through a function of the coincidence parameter $\tilde{r}$, and finally we also consider
vi)  two kinds of models with a self-interaction between DM, without DE. 
These models have been already studied in the literature and constrained with observational data available at that time. In order to discriminate between them at background level, we use the Statefinder diagnostic, based on the computation and study of the so-called Statefinder parameters $r$, $s$ in addition to 
the deceleration parameter $q$. We plot
the evolution trajectories for the several interacting models on the $r-q$, $r-s$ planes, and we find some distinctive features and departures from $\Lambda$CDM and other DE models, as Quintessence, Chaplygin Gas, running vacuum models (RVM) and Galileon.
\end{abstract}

\pacs{$\Lambda$CDM ; Interacting dark matter and dark energy ; Statefinder diagnostic.}

\maketitle

\section{Introduction}
\label{intro}
Since the advent of modern cosmology and the discoveries of the late twentieth century,
our view of the Universe has been firmly shifted. In this regard, 
current observational evidence coming from precise measurements of Supernovae Ia (SnIa) \cite{riess6071998} \cite{perlmutter1999measurements}, the large scale structure (LSS) from the Sloan Digital Sky Survey \cite{AdelmanMcCarthy:2007aa} and the cosmic microwave background (CMB) anisotropies \cite{Komatsu:2010fb} indicates that our Universe is flat or almost flat, expanding with an accelerated rate of expansion, and dominated by dark energy (DE) and cold dark matter (DM). 
The accelerating expansion of the present Universe is attributed to the DE, which is an exotic component with negative pressure, such as the cosmological constant (CC) $\Lambda$ \cite{Padmanabhan:2002ji,Sahni:1999gb,Carroll:2000fy}, which can be modeled as a perfect fluid with energy density $\rho_{\Lambda} = \Lambda/8\pi G$ and negative pressure $p_{\Lambda}=-\rho_{\Lambda}$ .The Lambda-cold-dark-matter ($\Lambda$CDM) model
has proven to be successful at explaining most observations
of our universe \cite{Bull:2015stt}. Despite this success, the $\Lambda$CDM model is plagued of theoretical issues. The first, and perhaps most significant problem, called the \emph{cosmological constant problem} \cite{Weinberg:1988cp,Martin:2012bt}, is the catastrophic disagreement between the value of the energy density of $\Lambda$, $\rho_{\Lambda}$, calculated from
observation, and the vacuum energy density, $\rho_{\textup{vac}}$, predicted by quantum field theory (QFT), which is about $10^{120}$ larger than the observed value. Since the theoretically expected value of the cosmological constant is much larger than the observed one, one may hope that $\Lambda$ vanishes in some way. In this case, an alternative explanation for the current accelerated expansion of the universe is required. Such explanations can be classified in modified matter models and modified gravity models. For a review of DE models, see Refs.\cite{Copeland:2006wr,Nojiri:2006ri,Joyce:2016vqv}.
The second significant problem with $\Lambda$ is the so-called \emph{coincidence problem} \cite{Zlatev:1998tr,Arkani-Hamed:2000ifx}. The coincidence parameter $\tilde{r}$ is defined as the ratio between DM and DE densities, $ \tilde{r} \equiv \frac{\rho_{c}}{\rho_{de}}$, which at the present time has a value of $\tilde{r_0} \sim  \mathcal{O}(1)$. In most models of the universe, this situation is indeed highly
coincidental, as a very specific set of initial conditions is required to yield the correct relative
energy densities in the present epoch. 

In addition to the already mentioned theoretical/conceptual issues, some anomalies and tensions have been found between cosmological and astrophysical data. In particular, the tension in estimation of the Hubble constant $H_0$ from different astronomical/cosmological observations assuming the standard $\Lambda$CDM model of the Universe has remained a big issue in the current cosmology \cite{DiValentino:2021izs}. A discrepancy between $4\sigma$ and $6\sigma$ \cite{Riess:2020fzl} in the measures of $H_0$ has been observed by comparing early measurements from Planck Collaboration \cite{Aghanim:2018eyx} and late observations with supernovae Type Ia \cite{Pan-STARRS1:2017jku}. In addition to the $H_0$ tension, a second tension has emerged during the recent years: the $S_8$ tension \cite{DiValentino:2020vvd}. In $\Lambda$CDM model, the $S_8$ parameter quantifies the amplitude of matter fluctuations in the late universe, and is defined as $(\Omega_{\mathrm matter}/0.3)^{0.5} \sigma_8$, where $\sigma_8$ is the standard deviation of the density fluctuation in an 8 $h^{-1}$ Mpc radius sphere. The measurement derived from low-redshift probes \cite{Planck:2016kqe,Battye:2014qga,Macaulay:2013swa} is systematically $2-3\,\sigma$  lower than that indirectly measured by the Planck-2018 CMB data \cite{Aghanim:2018eyx}. 
Theoretically, tensions between measurements in the early and the late universe could represent the signature of new physics beyond the $\Lambda$CDM model \cite{Mortsell:2018mfj}. On this subject, interacting DE models have been reexaminated in the light of latest cosmological observations, since a non-vanishing coupling between the dark sector of the universe can eventually alleviate 
the emerging cosmological tensions of $\Lambda$CDM model \cite{DiValentino:2017iww,Yang:2018euj,Yang:2018uae,Pan:2019gop,Sola:2019lnw,DiValentino:2019ffd,Gao:2021xnk,Hoerning:2023hks}. Remarkably, the assumption that DE and DM do not evolve separately but interact with each other non-gravitationally, which allows for an energy transfer between DE and DM, was considered previously in order to solve the cosmic coincidence problem \cite{Amendola:1999er, Amendola:2003eq, Amendola:2005ps, Pavon:2005yx, delCampo:2006vv, delCampo:2005tr, Olivares:2005tb, Olivares:2006jr, Wang:2005jx, Wang:2005ph, Wang:2006qw, Sadjadi:2006qp,Mangano:2002gg, Das:2005yj, Bean:2007nx, Bean:2007ny, Manera:2005ct, Nunes:2004wn, Chimento:2007yt, Valiviita:2008iv,delCampo:2008sr,Chimento:2009hj, Wei:2010fz, Arevalo:2011hh, Chimento:2012kg, Velten:2014nra, Grandon:2018hom, cosmoconstrain,Jesus:2020tby}. For extensive reviews on interacting DE models, see Refs.\cite{bolotin2015cosmological,Wang:2016lxa}. 
The coupling between DE and DM is modeled through a rate of energy exchange between both components. In the absence of a fundamental theory, this rate of energy exchange can not be obtained from first principles. Therefore, a phenomenological approach for the coupling is used \cite{Boehmer:2008av}. For a given model, it is possible to choose the parameters in order the model be consistent with data (e.g., SnIa, CMB, BAO and $H(z)$) that constraint the expansion history at background level \cite{Guo:2007zk,Valiviita:2009nu,Clemson:2011an,Salvatelli:2014zta,Nunes:2016dlj,Costa:2016tpb}. As several DE models predict very similar expansion histories, mostly of them are still in agreement with the available observational data. What is more, different mathematical model based on the same observational data could produce equivalent results \cite{sep-scientific-underdetermination}. Thus, we need a criterion to discriminate between DE models. They should agree with current observations and predict, at least, very similar expansion histories. The above-mentioned problem was one of the main reasons to introduce alternative mechanisms to identify models which could be quite similar than other. On this subject, the Statefinder diagnostic was introduced in 2003 by \cite{sahni2003statefinder,Alam:2003sc}. Such approach is a tool that makes possible to differentiate between the very distinct and competing cosmological scenarios involving DE at background level. The Statefinder diagnostics is performed taking advantage of higher order of derivatives on the scale factor $a(t)$, in such a way that it introduces the so-called Statefinder parameters, $r,s$, which are corrections to the Hubble rate $H(z)$ and deceleration parameter $q(z)$ of higher order in $a(t)$. Thus, by computing the correction behind $H(z)$ and $q(z)$ we could distinguish between models apparently equivalent. This diagnostic proposal introduces new geometrical dimensionless parameters that characterize the properties of DE regardless of the model, because they depend on the observable Hubble parameter and its derivatives \cite{granda2013natural}.
Statefinder diagnostic have been applied to several models, e.g., some authors have applied the Statefinder diagnostic to discriminate holographic DE (HED) models from the $\Lambda$CDM model \cite{Zhang:2005yz}, to analyze Barrow holographic DE \cite{Pradhan:2021cbj}, interacting DE models \cite{Zimdahl:2003wg,Cui:2015ueu,Jiang:2020pmt,Panotopoulos:2019xbw}, among others \cite{Panotopoulos:2007zn,Panotopoulos:2018sso,Sharma:2020mzl,Pacif:2021khn,Alvarez:2022mlf}.

Considering that an interaction between DE and DM offers an interesting framework to study phenomenology
beyond to $\Lambda$CDM model, the main goal of the present work is to analyse several interacting DE, recently proposed in the literature, via Statefinder diagnostic. In doing so, for each interacting DE model we compute the Statefinder parameters as functions of the redshift, studying their high and low-redshift limits. We also plot the evolution trajectory on the $s-r$ and $q-r$ planes in order to determine its deviation from $\Lambda$CDM. In addition, according to their behaviour on the $s-r$ and $q-r$ planes, we discriminate between the several interacting DE studied, breaking the degeneracies of the models at background level. Basically, we perform
the Statefinder diagnostic for 17 several forms of the  energy transfer rate $Q$ between DE and DM, belonging to the following categories:
i) linear models in energy densities of DE and DM, 
ii) non-linear models, 
iii) models with a change of direction of energy transfer between DE and DM, 
iv) models involving derivatives of the energy densities, 
v) parametrized interactions through a function of the coincidence parameter $\tilde{r}$, and finally we also consider two kinds of models with a self-interaction between DM, without DE. These models have been already studied in the literature and constrained with observational data available at that time, see e.g. \cite{He:2008tn,Chimento:2009hj,Wei:2010fz,hao2011cosmological,Arevalo:2011hh,Pan:2012ki,bolotin2015cosmological,Pan:2016ngu,Wang:2016lxa,Xia:2016vnp,Yang:2017zjs,DiValentino:2017iww,cosmoconstrain,Yang:2018euj,DiValentino:2019ffd,Arevalo:2022sne}.

Our work is organized as follows: after this introduction, we present a brief review of the $\Lambda$CDM model as well as the basic equations which describe an interaction between DE and DM in 
Section \ref{section2}. In the third Section, we present 
the Statefinder parameters which will be applied for the several models, while in section \ref{section4} the background dynamics of each particular model is studied. In particular, analytical solutions for the corresponding Hubble rate as functions  of the redshift $H(z)$ are presented. In the fifth section we discuss the Statefinder analysis for the several models, and in Section \ref{section6} we make the comparison between models in light of the Statefinder analysis. Finally we summarize our findings and present our conclusions in Section \ref{section7}. We adopt the mostly positive metric signature, $(-,+,+,+)$, and we work in natural units where $c=\hbar=1$.\\

\section{Theoretical Framework}\label{section2}

\subsection{$\Lambda$CDM model}

We consider a FLRW Universe
\begin{equation}
ds^2 = -dt^2 + a(t)^2\left[\frac{dr^2}{1-K r^2}+r^2 d\theta^2+r^2\sin^2 \theta d\phi^2\right],
\end{equation}
where $a(t)$ is the scale factor at cosmic time t and $K$ is the Gaussian curvature of the space-time. Setting $\kappa^2 = 8 \pi G$, with $G$ being the Newton's constant, the scale factor satisfies the Friedmann equations \cite{dodelson2021modern}
\begin{eqnarray}
H^2 +\frac{K}{a^2}  & = & \frac{\kappa^2}{3}\sum_{A}\rho_A+\frac{\Lambda}{3}, \label{F1} \\
\frac{\ddot{a}}{a} & = & -\frac{\kappa^2}{6}\sum_{A}(\rho_A + 3 p_A)
+\frac{\Lambda}{3} \label{f2},
\end{eqnarray}
where $H=\dot{a}/a$ is the Hubble rate and $\rho_A$ and $p_A$ denote the energy density and pressure of each individual fluid component, respectively, and they are related by an en equation-of-state (EoS) parameter $w_A=p_A/\rho_A$. The curvature term can be brought to the right-hand side of Eq.(\ref{F1}) by defining $\rho_K=-3K/(8\pi G a^2)$. On the other hand, the energy density associated to $\Lambda$ can be written as $\rho_{\Lambda} = \Lambda/8\pi G$, while its EoS becomes $w_{\Lambda}=-1$. In addition to the curvature and cosmological constant terms, the $\Lambda$CDM model also includes other energy density components: baryons (b), cold DM (c), and radiation (r), characterized by the EoS parameters $w_b=w_c=0$ and $w_r=1/3$, respectively.

In $\Lambda$CDM it is assumed that the different matter components do not have any other interaction, therefore the energy conservation equation for each component holds
\begin{equation}
    \dot{\rho}_{A}+3H(1+w_A)\rho_{A}=0.\label{cons}
\end{equation}
Then, by integrating Eq.(\ref{cons}) for baryons, cold DM, and radiation, the Friedmann equation for $\Lambda$CDM (\ref{F1}) can be written as
\begin{equation}\label{f11}
    H^2=H^2_0 \left[\frac{\Omega_{r0}}{a^4} +\frac{\Omega_{m0}}{a^3}+\frac{\Omega_{K0}}{a^2}+\Omega_{\Lambda} \right],
\end{equation}
where we have used the fact that each energy density is usually expressed in terms the dimensionless density parameter, defined as $\Omega_A=\rho_A/\rho_{cr}$ with $\rho_{cr}=3H^2/(8\pi G)$ being the critical density, and the sub-index “0” denotes the current value of any given quantity. The Hubble constant is usually expressed as $H_0=100 h\,\mathrm{km~s^{-1}~Mpc^{-1}}$, where the parameter $h$ denotes the observational uncertainty. In Eq.(\ref{f11}), $\Omega_{m0}$ denotes the current contribution of non-relativistic matter (baryons+cold DM). By convention, the dimensionless density parameter associated with the curvature term is $ \Omega_{K}=-K/(aH)^2$ \cite{liddle2015introduction, piattella2018lecture}.

The cosmological parameters of the $\Lambda$CDM model, derived from the CMB temperature fluctuations measured by Planck Collaboration with the addition of external data sets are given by \cite{Lahav:2022poa}

\begin{align*}
    h&=0.674 \pm 0.005,
    \hspace{1cm}
    \Omega_{m0}=0.315 \pm 0.007,
    \hspace{1cm}
    \Omega_{c0}h^2=0.1200 \pm 0.0012, 
    \\
    \Omega_{b0}h^2&=0.02237 \pm 0.00015,
     \hspace{0.4cm}
    \Omega_{\Lambda0}=0.685 \pm 0.007,
     \hspace{1cm}
    \Omega_{m0}=0.315 \pm \pm 0.007,
\end{align*}
where the spatial flatness ($K=0$) has been assumed, while the contribution of radiation becomes $\Omega_{r0}\sim 10^{-5}$.

By using the relation between the scalar factor and the redshift $a=(1+z)^{-1}$, the Friedmann equation (\ref{f11}) for $\Lambda$CDM with zero curvature may be written as
\begin{equation}\label{LCDM}
   {E^2(z)=\Omega_{r0}(1+z)^{4}+\Omega_{b0}(1+z)^{3}+\Omega_{c0}(1+z)^{3}+\Omega_{\Lambda0},}
\end{equation}

where $E(z)\equiv H(z)/H_0$ is defined as the dimensionless Hubble rate. At $z=0$ (today), the following constraint is satisfied
\begin{equation}\label{cons}
    \Omega_{r0}+\Omega_{b0}+\Omega_{c0}+\Omega_{\Lambda0}=1.
\end{equation}

If we set $\Lambda=0$ and consider a more general DE (de) component as the responsible of the accelerated expansion, having energy density $\rho_{de}$ and EoS $w_{de}$, a combination of CMB, SN, and BAO measurements, assuming a flat Universe, found $w_{de}=-1.03\pm 0.03$ \cite{Aghanim:2018eyx}, consistent with the cosmological constant case $w_{\Lambda}=-1$.

\subsection{Interaction between DE and DM}

Now, we are going to consider the case where an interaction between a dynamical DE component (with and EoS parameter $w_{de}$) and DM is allowed. Despite the additional postulated interaction, the total energy density of the dark sector is conserved, and the energy momentum tensor of the dark sector,$T_{\nu}^{\mu}=T_{(de)_{\nu}}^{\mu}+T_{(c)_{\nu}}^{\mu}$, satisfies
\begin{equation}
    \triangledown_{\mu}T_{\nu}^\mu=\triangledown_{\mu}(T_{(de)_{\nu}}^{\mu}+T_{(c)_{\nu}}^{\mu})=0,
\end{equation}
therefore, if an interchange of energy between DE and DM is introduced, according to \cite{Valiviita:2009nu,bolotin2015cosmological}. 
\begin{equation}\label{int1}
    \triangledown_{\mu}T_{(de)_{\nu}}^{\mu}=-\triangledown_{\mu}T_{(c)_{\nu}}^{\mu}=F_\nu,
\end{equation}
where the $F_\nu$ is the four-vector of interaction between dark components.
We can project the Eqs. (\ref{int1}) parallel to the four-velocity $u^{\mu}$, and we obtain
\begin{equation}\label{par}
    u^{\mu}\triangledown^{\nu}T_{(c)_{\mu\nu}}=-u^{\mu}F_\mu,~~~~~~
    u^{\mu}\triangledown^{\nu}T_{(de)_{\mu\nu}}=u^{\mu}F_\mu,
\end{equation}
and respect part orthogonal to the velocity, we must use the projector $h_{\beta\mu}=g_{\beta\mu}-u_\beta u_\mu$ 
\begin{equation}\label{per}
    h^{\mu\beta}\triangledown^{\nu}T_{(c)_{\mu\nu}}=-h^{\mu\beta}F_\mu,~~~~~~
    h^{\mu\beta}\triangledown^{\nu}T_{(de)_{\mu\nu}}=h^{\mu\beta}F_\mu,
\end{equation}
In addition, the general expression for the energy-momentum tensor tensor describing perfect fluid is \cite{carrollspacetime}, \cite{piattella2018lecture}
\begin{equation}\label{em}
    T^{\mu\nu}=(\rho+p)u^\mu u^\nu-pg^{\mu\nu}.
\end{equation}
Now, we use Eqs. (\ref{em}), (\ref{par}) and (\ref{per}) to obtain the following Euler equations 
\begin{equation}\label{euler1}
h^{\mu\beta}\triangledown_{\mu}p_c+(\rho_c+p_c)u^{\mu}\triangledown_{\mu}u^{\beta}=-h^{\mu\beta}F_{\mu},
\end{equation}
\begin{equation}\label{euler2}
h^{\mu\beta}\triangledown_{\mu}p_{de}+(\rho_{de}+p_{de})u^{\mu}\triangledown_{\mu}u^{\beta}=h^{\mu\beta}F_{\mu}.
\end{equation}
In the context of a flat FLRW universe, $u^{\mu}=(1,0,0,0)$ in the comoving coordinates, thus
\begin{align}
    \triangledown_{\mu}u^{\mu}&=3H,\\
    u^{\mu}\triangledown_{\mu}u^{\nu}&=0.
\end{align}
We use the notation $u^{\mu}F_{\mu}=Q$. Now, we write the Eqs. (\ref{euler1}) and (\ref{euler2}) in the form
\begin{align}
    \dot{\rho_{c}}+3H\rho_c&=Q,\\
    \dot{\rho}_{de}+3H(\rho_{de}+p_{de})&=-Q,
\end{align}
Where $Q$ is the rate of energy exchange between DE and DM. 
Here, $Q>0$ reflects that the energy flows from DE to DM, while $Q<0$ the opposite. In a spatially flat FLRW universe filled with DE, DM, baryons, and radiation, we can split the equation (\ref{cons}) for baryons and radiation, yielding the following set of equations
\begin{equation}\label{q1}
\dot{\rho}_{de}+3H(1+w_{de})\rho_{de}=-Q,
\end{equation}
\begin{equation}\label{q2}
\dot{\rho}_{c}+3H\rho_{c}=Q,
\end{equation}
\begin{equation}\label{bar}
\dot{\rho}_{b}+3H\rho_{b}=0,
\end{equation}
\begin{equation}\label{rad}
\dot{\rho}_{r}+4H\rho_{r}=0,
\end{equation}
In phenomenological construction of interacting DE models, it is assumed that
$Q$ can be modeled as a certain function of energy densities and the Hubble rate \cite{bolotin2015cosmological,Wang:2016lxa}. Several expressions for the rate of energy exchange $Q$ have been studied in the literature: models where $Q$ is a linear function (widely studied) of densities of DM or DE \cite{DiValentino:2019ffd,Yang:2018euj,Yang:2017zjs,Pan:2016ngu,Xia:2016vnp,Chimento:2009hj,He:2008tn}, nonlinear interactions \cite{Yang:2017zjs,cosmoconstrain,bolotin2015cosmological,Chimento:2009hj}, proportional to deceleration parameter \cite{hao2011cosmological}, and others combinations, where $Q$ is constructed from derivatives of the densities of DM or DE, or varying couplings with dependence on cosmic time or redshift \cite{He:2008tn}, among others.
\\
The conservation equations for baryons and radiation, Eqs.(\ref{bar}) and (\ref{rad}), respectively, are solved by separation of variables, yielding
\begin{equation}\label{bar2}
\rho_b(z)=\rho_{b0}(1+z)^3,
\end{equation}
\begin{equation}\label{rad2}
\rho_r(z)=\rho_{r0}(1+z)^4.
\end{equation}
However, the solutions of the equations (\ref{q1}) and (\ref{q2}) depend on the mathematical structure of $Q$, therefore, it is not always possible to obtain an analytical solution for $E(z)=H(z)/H_0$. Then, we focus on interacting DE models that allow analytical solutions for $E(z)$ in order to apply the Statefinder analysis. 
\bigskip

\section{Statefinder Diagnostic}\label{section3}

\begin{figure}[htp]
  \centering
  \label{figurL}
  \subfloat[]{\label{figurL:1}\includegraphics[width=60mm]{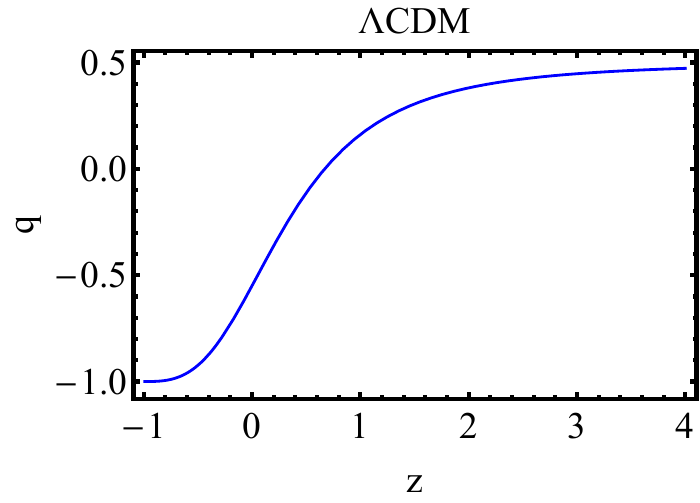}}
  \subfloat[]{\label{figurL:2}\includegraphics[width=60mm]{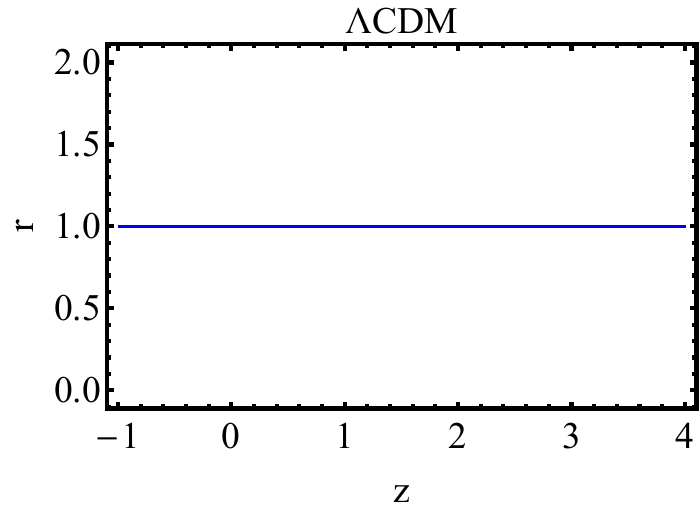}}
  \\
  \subfloat[]{\label{figurL:3}\includegraphics[width=60mm]{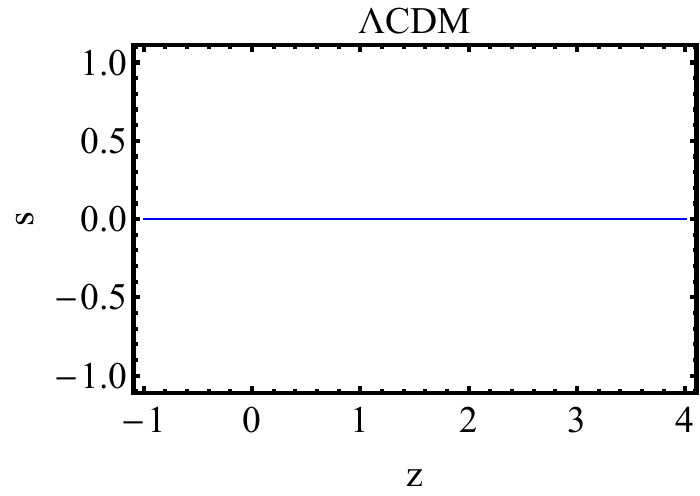}}
    \\  
  \subfloat[]{\label{figurL:4}\includegraphics[width=60mm]{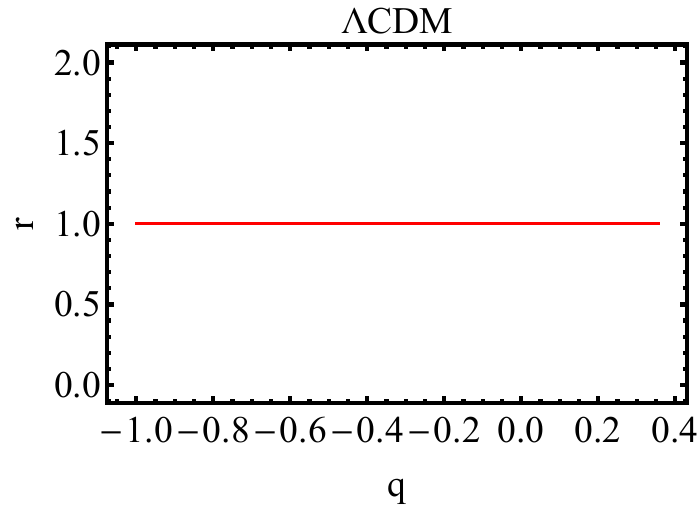}}
  \subfloat[]{\label{figurL:5}\includegraphics[width=60mm]{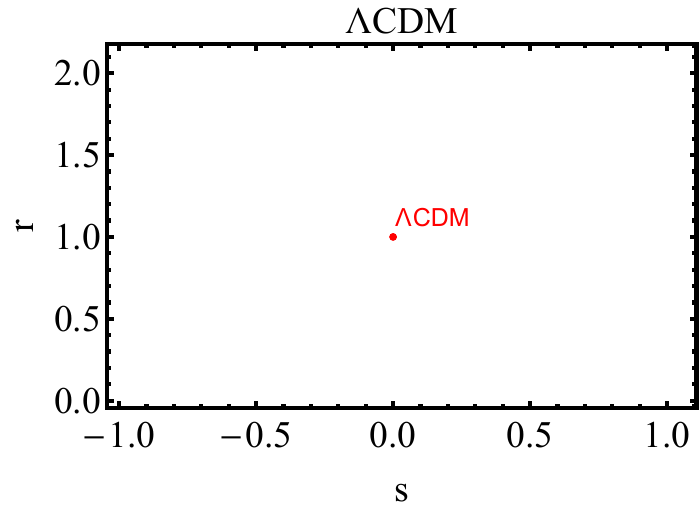}}
 \caption{Statefinder parameters plots $q(z),r(z),s(z)$, $q-r$  and $s-r$ for $\Lambda$CDM.}
\end{figure}

In Ref.\cite{Alam:2003sc}, the authors introduced the Statefinder parameters by expanding in Taylor series the scale factor $a(t)$. We are interested in small values of $|t-t_0|$. The scale factor $a(t)$ is expanded as follows
\begin{equation}
a(t) = a(t_0)+ \dot{a}|_0(t-t_0)+ \frac{\ddot{a}|_0}{2}(t-t_0)^2+\frac{\dddot{a}|_0}{6}(t-t_0)^3+...,
\end{equation}
or in the form \cite{Arabsalmani:2011fz}
\begin{equation}
(1+z)^{-1}:=\frac{a(t)}{a(t_0)}=1 + \sum_n \frac{A_n(t_0)}{n!} [H_0(t-t_0)]^n,
\end{equation}
where $A_n=\frac{a^{(n)}}{aH^n}, n\in \mathbb{N}, a^{(n)}$ is $n$th derivative of the scale factor with respect to cosmic time. $A_n$ is used to define the Statefinder parameters $q$, $r$ and $s$. Recalling that $q$ is defined above as the deceleration parameter, it can be written in terms of $A_2$ as
\begin{equation}
q \equiv -A_2=-\frac{\ddot{a}}{a{H^2}}=-1-\frac{\dot{H}}{H^2}.
\end{equation}
We rewrite  Eq. (\ref{f2}) in terms of the deceleration parameter as follows
\begin{equation}
  q= -\frac{\ddot{a}}{a{H^2}}=\sum_{i}^{} \frac{4\pi G \rho_i}{3H^2}(1+3w_i)=\frac{(1+3w_{de}\Omega_{de})}{2},
\end{equation}
where $w_i=p_i/\rho_i$, $\Omega_i=8\pi G \rho_i/3H_0^2$ and we have assumed that $\Omega_b+\Omega_c+\Omega_{de}=1$.
The parameter $r$ is the next (after the Hubble rate $H$ and the deceleration parameter $q$) member of the set of kinematic parameters that describe the expansion of the Universe \cite{cosmoevo}, which is defined as
\begin{equation}
r\equiv A_3=\frac{\dddot{a}}{aH^3},
\end{equation}
or in terms of DE density 
\begin{equation}\label{rrr}
r=1+\frac{9}{2}\Omega_{de}w_{de}(1+w_{de})-\frac{3}{2}\Omega_{de}\frac{\dot{w}_{de}}{H},
\end{equation}
where the parameter $r$ was first introduced in \cite{Chiba:1998tc}, and in \cite{Blandford:2004ah} as well as the \textit{jerk} ‘$j$’.\\
On the other hand, the parameter $s$ is defined as a combination of $r$ and deceleration parameter, and it should not to be confused with the \textit{snap} (the fourth time derivative) \cite{Visser:2003vq}
\begin{equation}\label{sss}
s\equiv \frac{r-1}{3(q-1/2)}, 
\end{equation}
substituting $q$ and $r$,
\begin{equation}
  s=\frac{r-1}{3(q-1/2)}=1+w_{de}-\frac{1}{3}\frac{\dot{w}_{de}}{w_{de}H}.  
\end{equation}
It is observed that $s$ does not explicitly depend on the DE density. The reason of why $s$ is chosen in such a way is due to the fact that the features chosen for the description of DE may be geometrical, if they come directly from the metric of space-time, but also physical, if they depend on the characteristics of the fields that represent DE. Physical qualities vary on models, while geometrical characteristics are universal \cite{cosmoevo}.
\\
For our analysis, it will be useful to express these parameters in terms of the dimensionless Hubble rate and the redshift $z$. The cosmic times derivatives are written as redshift derivatives according to
\begin{equation}\nonumber
\frac{d}{dt}=\frac{d}{dz}\frac{dz}{da}\frac{da}{dt}=-(1+z)H(z)\frac{d}{dz},
\end{equation}
where $E^2(z)=H^2(z)/H_{0}^2$ as been used. In this way, the set Statefinder parameters becomes

\begin{align}\label{ese}
    q&=-1-\frac{1+z}{2E^2}\frac{dE^2}{dz},\\
    r&=1+\frac{1}{2E^2}\left[(1+z)^2\frac{d^2E^2}{dz^2}-2(1+z)\frac{dE^2}{dz}\right],\\ 
    s&=\frac{1}{3} \frac{(1+z)^2\frac{d^2E^2}{dz^2}-2(1+z)\frac{dE^2}{dz}}{(1+z)\frac{dE^2}{dz}-3E^2}.
\end{align}
Therefore, if the role of DE is played by a cosmological constant, i.e $w_{de} = -1$, then the value of $r$ remains at $r =1$ throughout the matter-dominated epoch and at all future times (for $z \lesssim 10^4$). In the $s-r$ plane, the fixed point $\{s,r\}= \{0,1\}$ in a Universe containing a cosmological constant and non-relativistic matter corresponds to $\Lambda$CDM.\\
According to \cite{Alam:2003sc, Sahni:DMDE}, a first limit case ($\Omega_m =\Omega_c+\Omega_b= 1$, $\Omega_{de}= 0$), without radiation, corresponds to the standard cold dark matter (SCDM), in which the Universe presents a decelerated power-law expansion, according to $a(t) \propto t^{2/3}$. In this case, the fixed point is found to be $\{s,r\} = \{1,1\}$. The other limit case ($\Omega_m =0$, $\Omega_{de}= 1$) corresponds to a dS Universe,  which expands at constant rate, $a(t) \propto \mathrm{exp} \sqrt{\frac{\Lambda}{3}}t$.\\
The following plots correspond to the case for $\Lambda$CDM model. In Figs. (\ref{figurL:1}), (\ref{figurL:2}), (\ref{figurL:3}) we show the plots of Statefinder parameters against redshift $q(z),r(z)~~ \mathrm{and}~ s(z)$, and in Figs. (\ref{figurL:4}), (\ref{figurL:5}) are shown the parametric plots in $s-r$ and $q-r$ planes. In addition, the fixed point $\{q,r\} = \{-1,1\}$ in the $q - r$ plane corresponds to the asymptotic dS solution \cite{granda2013natural}. \\
By comparing the evolution trajectories on the $r-q$ and $r-s$ planes 
between the corresponding to several DE and $\Lambda$CDM models, it is 
possible to find some similarities and analyze the deviation and compatibility with $\Lambda$CDM model.

\section{Interacting DE models: analytic solutions for $H(z)$}\label{section4}

In this Section, we present several mathematical expressions for the coupling $Q$ between DE and DM as well as the corresponding solutions for the dimensionless Hubble rate $E=H(z)/H_0$, which will be used in order to analyze each model by means the Statefinder diagnostic.
First, we present models 1, 2, where $Q$ is linear with respect to the energy densities of DM and DE, respectively. Model 3 is a linear combination of DE and DM energy densities. Models 4, 5, and 6 are non-linear. Models 7, 8, and 9 involve a coupling proportional to decelerating parameter $q$. Model 10 considers a linear combination of the derivatives (respect to ln$(a^3)$) of the energy densities DM and DE. Furthermore, models 11, 12, 13, 14, and 15 are interactions that can be parameterized as a function of the ratio $\tilde{r}$ between density DM and DE. Finally, as an "extra bonus", models 16 and 17 are two kinds of special models without DE, where exists a self-interaction between DM. It should be noted that all the investigated models are dimensionally consistent with respect to the expression for $Q$. In this paper, all the expressions for $Q$ we focused on may written in general terms as $Q=3H\gamma R(\rho_c, \rho_{de})$, where $\gamma$ is a dimensionless constant and $R$ is a real function having units of energy density. Moreover, each $Q$ is linear in the Hubble rate $H$, since this facilitates the consistency of units and the solving of differential equations by replacing $Q$ in Eqs. (\ref{q1}) and (\ref{q2}). On the other hand, we will assume $w=const.$ where it corresponds.

\subsection{Linear interactions}

Going further the simplest interacting DE models where $Q$ is constant, linear models have brought a lot of interest in their study. Accordingly, we are going to study three linear models, $Q$ proportional to the DM density, the second proportional to the DE, and a general case being a linear combination of DE and DM energy densities. The models with the rate of energy transfer $Q \propto \rho_c$ and $Q \propto \rho_{de}$ have been widely studied in the literature. We use the expressions presented in Refs. \cite{Xia:2016vnp}, \cite{Grandon:2018hom}, where $Q$is proportional to the constant $3\gamma$, since this simplifies the expression for $E^2(z)$. 

\subsubsection{MODEL 1. $Q_1=3\gamma H \rho_c$}

For the first case to be presented, the rate $Q$ is proportional to the DM energy density
\begin{equation}
    Q=3\gamma H \rho_c,\label{Q1}
\end{equation}
where $\gamma$ is a dimensionless constant and gives us the strength of the coupling. If we set $K=0$, the Friedmann equation (\ref{F1}) becomes
\begin{equation}\label{fri}
    H^2=\frac{8\pi G}{3}[\rho_r+\rho_b+\rho_c+\rho_{de}].
\end{equation}
Assuming that $w_{de}$ is a constant, the authors in Refs.\cite{Xia:2016vnp,Chimento:2012kg}, provide the analytical solutions for the densities $\rho_c$ and $\rho_{de}$ by inserting Eq.(\ref{Q1}) into Eqs. (\ref{q1}) and (\ref{q2}), thus
\begin{align}
    \rho_c&=\rho_{c0}(1+z)^{3(1-\gamma)},\\
    \rho_{de}&=\left[\rho_{de,0}+\rho_{c0}\frac{\gamma}{w_{de}+\gamma}\right](1+z)^{3(1+w_{de})}.
\end{align}
If we substitute Eqs. (\ref{bar2}), (\ref{rad2}), the expression for $\rho_c$, $\rho_{de}$ and $a=(1+z)^{-1}$ into (\ref{fri}), and dividing by the critical density, the expression for the dimensionless Hubble rate $E(z)=H(z)/H_0$ is expressed as
\begin{equation}
E^2(z)=\Omega_{de}(1+z)^{3(1+w_{de})}+\Omega_r(1+z)^4+\Omega_b(1+z)^3+\Omega_{c}\left(\frac{\gamma}{w_{de}+\gamma}(1+z)^{3(1+w_{de})}+\frac{w_{de}}{w_{de}+\gamma}(1+z)^{3(1-\gamma)}\right).
\end{equation}
Besides, the function $E(z)$ for $w_{de}=-\gamma$ is a special case to consider. We calculate the limit for that purpose. When $w \rightarrow -\gamma$ the expression for the dimensionless Hubble rate becomes
\begin{equation}
E^2(z) \rightarrow (1+z)^3\left[\Omega_{b}+\Omega_{r}+z\Omega_{r}-(1+z)^{-3\gamma }(\Omega_{b}+\Omega_{r}-3 \gamma \Omega_{c} \log(1+z)-1))\right].
\end{equation}

\subsubsection{MODEL 2. $Q_2=3\gamma H \rho_{de}$}
The second interaction within this category depends linearly on the DE energy density
\begin{equation}
  Q=3\gamma H \rho_{de}.\label{Q2}
\end{equation}
In the same way as the first interaction, according to Refs. \cite{Chimento:2012kg}, \cite{Xia:2016vnp}, \cite{Grandon:2018hom}, if Eq.(\ref{Q1}) is replaced into Eq.(\ref{q1}), the solution for $\rho_{de}$ is found to be
\begin{equation}
    \rho_{de}=\rho_{de,0}(1+z)^{3(1+w_{de}+\gamma)}.
\end{equation}
Accordingly, the expression for the dimensionless Hubble rate yields
\begin{equation}\label{m2}
E^2(z)=\Omega_m(1+z)^3+\Omega_r(1+z)^4+\Omega_{de}\left(\frac{\gamma}{w_{de}+\gamma}(1+z)^3+\frac{w_{de}}{w_{de}+\gamma}(1+z)^{3(1+w_{de}+\gamma)}\right),
\end{equation}
where $\Omega_m=\Omega_c+\Omega_b$,~$\Omega_r=2.469~\times~10^{-5}h^{-2}(1+0.2271N_{eff})$ and $N_{eff}=3.046$. \\
If $\gamma<0$, this implies that the energy transfer is from DM to DE, and for MODELS 1 and 2 if we set $\gamma=0$, there is no interaction and both models reduced to $w$CDM, while for $w_{de} =-1$, $w$CDM reduces to Eq. (\ref{LCDM}) corresponding to $\Lambda$CDM. \\
Analogously to the first case, if $w_{de} \rightarrow -\gamma$, the expression for the dimensionless Hubble (\ref{m2}) takes the form 
\begin{equation}
E^2(z) \rightarrow (z+1)^3[1+z\Omega_{r}-3\gamma\Omega_{de}\log (z+1)].
\end{equation}

%%%%%%%%%%%%%%%%%%%%%%%%%%%%%%%%%%%%%%%%%%%%%%%%%%%%%%%%%%%%%%%%%%%%%%%%%%%%%%%%
\subsubsection{MODEL 3. $Q_3= 3H(\lambda_{c}\rho_{c} +\lambda_{de}\rho_{de})$}
%%%%%%%%%%%%%%%%%%%%%%%%%%%%%%%%%%%%%%%%%%%%%%%%%%%%%%%%%%%%%%%%%%%%%%%%%%%%%%%%

Our third model is a linear combination of DE and DM energy densities
\begin{equation}
    Q= 3H(\lambda_{c}\rho_{c} +\lambda_{de}\rho_{de})\label{Q3},
\end{equation}
which has been studied in Refs.\cite{Quartin:2008px,He:2008tn,Chimento:2009hj,Chimento:2012kg,Pan:2012ki,Pan:2016ngu}. 
The coupling parameters $\lambda_c$ and $\lambda_{de}$ denote the strengths of the interaction between the components of the dark sector and their signs imply the direction of energy transfer. It is relevant to mention that in \cite{Pan:2012ki}, the analytic solution found for this interaction assumed that the coupling parameters are very small (i.e. $\left| \lambda_c \right| \ll 1$ and 
$\left| \lambda_{de} \right| \ll 1$ ), and the authors neglect the product $\lambda_c \lambda_{de}$ in the solution, however, we will consider this product to be not null when using the expressions in the subsequent analysis.
In Ref.\cite{Pan:2016ngu}, the conservation equations (\ref{q1},(\ref{q2}) were unified and the authors derived the following master equation when the coupling (\ref{Q3}) is considered
\begin{equation}\label{master}
    \rho_T'' + 
    \left(2+w_{de}+\lambda_{de}-\lambda_{c}-\frac{w_{de}'}{w_{de}}\right)\rho_T' +\left[(1+w_{de})(1-\lambda_c)+\lambda_{de}-\frac{w_{de}'}{w_{de}}\right]\rho_T=0,
\end{equation}
which is a second order differential equation, where $\rho_T = \rho_{c}+\rho_{de}$ and primes denote derivatives respect to the variable $x=\ln(a^3)$.
In order to solve the above equation, for $w_{de}=const.$, the solution of Eq. (\ref{master}) according to Refs. \cite{Chimento:2009hj}, \cite{Pan:2016ngu} is found to be
\begin{equation}
    \rho_T=\rho_1a^{3m_+}+\rho_2a^{3m_-},
\end{equation}
where $\rho_1,\rho_2$ are integration constants, and $m_{\pm}$ are
\begin{equation}
m_{\pm} = \frac{\lambda_c - w_{de} - \lambda_{de} - 2 \pm
    \sqrt{(\lambda_c + w_{de} + \lambda_{de})^2 - 4 \lambda_c\lambda_{de}}}{2}, 
\end{equation}
considering $(\lambda_c + w_{de} + \lambda_{de})^2 - 4 \lambda_c\lambda_{de}>0$ to obtain $m_{+}$ and $m_{-}$ real and distinct. From Eq. (\ref{F1}) and neglecting the contribution of radiation, the Hubble rate takes the following form
\begin{equation}\label{hl1}
    H^2=\frac{8\pi G}{3}\left[\rho_{b0}a^{-3}+\rho_1a^{3m_{+}}+\rho_2a^{3m_{-}}\right]-\frac{K}{a^2}.
\end{equation}
The explicit analytic solutions for DM and DE energy densities are
\begin{eqnarray}
    \rho_{c} &=& \rho_{1}\frac{w_{de}+1+m_{+}}{w_{de}}a^{3m_{+}}+\rho_{2}\frac{w_{de}+1+m_{-}}{w_{de}}a^{3m_{-}}, \\
    \rho_{de} &=& -\frac{\rho_{1}(1+m_{+})a^{3m_{+}}+\rho_{2}(1+m_{-})a^{3m_{-}}}{w_{de}},
 \end{eqnarray}
respectively. Furthermore, in terms of the density parameters for the equivalent two fluids $\Omega_1$ and $\Omega_2$, the density parameters for DM ($\Omega_{c0}$) and DE ($\Omega_{de,0}$) are respectively written as
\begin{equation}\label{omc}
\Omega_{c0}=\Omega_1\frac{(w_{de}+1+m_{+})}{w_{de}}+\Omega_2\frac{(w_{de}+1+m_{-})}{w_{de}},
\end{equation}
\begin{equation}\label{omde}
\Omega_{de,0}=-\frac{\Omega_1(1+m_{+})+\Omega_2(1+m_{-})}{w_{de}},
\end{equation}
where $\Omega_i=8\pi G\rho_i/3H_0^2$. Using the aforementioned equations (\ref{omc}) and (\ref{omde}), the parameters $\Omega_1$ and $\Omega_2$ may be expressed. Taking into account $\Omega_{c0}+\Omega_{de,0}=\Omega_1+\Omega_2$, and 
\begin{equation}
\Omega_{c0} + \Omega_{de,0} + \Omega_{b0} + \Omega_{K} = \Omega_{1} + \Omega_{2} + \Omega_{b0} + \Omega_{K} = 1, 
\end{equation}
from Friedmann equation (\ref{F1}) at the present time $t=t_0$. Then, it is obtained 
\begin{equation}\nonumber
\Omega_{1} = \frac{w_{de}\Omega_{c0}-(1+w_{de}+m_{-})(1-\Omega_{b0}-\Omega_{k})}{m_{+}-m_{-}}. 
\end{equation}
For a flat Universe, i.e. $\Omega_K=0$, we have that
\begin{equation}\label{oc}
\Omega_c=\Omega_1\frac{(w_{de}+1+m_{+})}{w_{de}}a^{3m_{+}}+(1-\Omega_1-\Omega_{b0})\frac{(w_{de}+1+m_{-})}{w_{de}}a^{3m_{-}},
\end{equation}
\begin{equation}\label{ode}
\Omega_{de}=-\frac{\Omega_1(1+m_{+})a^{3m_{+}}+(1-\Omega_1-\Omega_{b0})(1+m_{-})a^{3m_{-}}}{w_{de}}.
\end{equation}
If we write Eq. (\ref{hl1}) in terms of the normalized densities and replace  $\Omega_c$, $\Omega_{de}$ and $a=(1+z)^{-1}$, the dimensionless Hubble rate is expressed as follows
\begin{equation}
E^2(z)=\Omega_{b0}(1+z)^3+\Omega_{c}+\Omega_{de},
\end{equation}
and expanding it, we get
\begin{equation}
E^2(z)=\Omega_{b0}(1+z)^3+\Omega_1(1+z)^{-3m_{+}}
+(1-\Omega_1-\Omega_{b0})(1+z)^{-3m_{-}}.
\end{equation}
It is straightforward to check that if we set $\lambda_c=\lambda_{de}=0$ and $w_{de}= -1$, there is no interaction and the model reduces to $\Lambda$CDM.

\subsection{Non-linear interactions}

The next three non-linear interactions we have already been studied in \cite{Arevalo:2011hh,Chimento:2012kg,bolotin2015cosmological}. First, we recall that the coincidence parameter $\tilde{r}$ is defined as the ratio between DM and DE energy densities
\begin{equation}
    \tilde{r} = \frac{\rho_{c}}{\rho_{de}}\label{rtilde}.
\end{equation}
If $\rho=\rho_{c}+\rho_{de}$, the individual energy densities are written as
\begin{equation}
    \rho_{c}=\frac{\tilde{r}}{1+\tilde{r}}\rho,~~~~~~\rho_{de}=\frac{1}{1+\tilde{r}}\rho.
\end{equation}
This allows us to obtain a differential equation for $r$, which makes it easier to find its solution, as we will show later in the parameterized interactions.
\\
\subsubsection{MODEL 4. $Q_4= 3H \gamma \frac{\rho_{c}\rho_{de}}{\rho}$}

For the first non-linear interaction
\begin{equation}
    Q_4= 3H \gamma \frac{\rho_{c}\rho_{de}}{\rho}\label{Q4},
\end{equation}
where $\rho=\rho_{c}+\rho_{de}$. According to \cite{Arevalo:2011hh}, the following analytic solutions were found
\begin{align}
    \tilde{r}&=\tilde{r}_0a^{3(w_{de}+\gamma)},\\
    \rho&=\rho_0a^{-3(1+w_{de})}\left[\frac{1+\tilde{r}_0a^{3(w_{de}+\gamma)}}{1+\tilde{r}_0}\right]^\frac{w_{de}}{w_{de}\gamma},\\
    \rho_{c}&=\rho_{c0}a^{-3(1-\gamma)}\left[\frac{1+\tilde{r}_0a^{3(w_{de}+\gamma)}}{1+\tilde{r}_0}\right]^{-\frac{\gamma}{w_{de}+\gamma}},\\
    \rho_{de}&=\rho_{de0}a^{-3(1+w_{de})}\left[\frac{1+\tilde{r}_0a^{3(w_{de}+\gamma)}}{1+\tilde{r}_0}\right]^{-\frac{\gamma}{w_{de}+\gamma}}.
\end{align}
Now, we construct the expression for the dimensionless Hubble rate from the Friedmann equation
\begin{equation}\label{321}
\begin{aligned}
    E^2(z)=\Omega_{de,0} \left(\frac{1}{1+z}\right)^{-3(w_{de}+1)} \left(\frac{\Omega_{de,0}+\Omega_{c0}
    \left(\frac{1}{1+z}\right)^{3(w_{de}+\gamma)}}{\Omega_{c0}+\Omega_{de,0}}\right)^{-\frac{\gamma}{\gamma +w_{de}}} \\
    + \Omega_{c0} \left(\frac{1}{1+z}\right)^{-3+3\gamma} \left(\frac{\Omega_{de,0}+\Omega_{c0}
    \left(\frac{1}{1+z}\right)^{3(w_{de}+\gamma)}}{\Omega_{c0}+\Omega_{de,0}}\right)^{-\frac{\gamma}{\gamma +w_{de}}} \\
    +\Omega_{b0}(1+z)^3+\Omega_{r0}(1+z)^4.
\end{aligned}
\end{equation}
$\Lambda$CDM is recovered for $\gamma=0$ and $w_{de}= -1$. For $z\gg 1$ (matter-dominated epoch), we obtain the expected evolution of density $\rho \propto a^{-3}$.\\
In the asymptotic case, if $w_{de} \rightarrow -\gamma$, the above expression (\ref{321}) becomes
\begin{equation}
\begin{aligned}
    E^2(z)\rightarrow (1+z)^3 \left[\Omega_{b0}+\Omega_{r0}+z\Omega_{r0}-(\Omega_{b0}+\Omega_{r0}-1) \left(1+z\right)^{-3 \gamma  \left[\frac{\Omega_{c0}}{\Omega_{b0}+\Omega_{r0}-1}+1\right]}\right],
   \end{aligned}
\end{equation}
and when $\gamma=1$, we have 
\begin{equation}\label{4red}
\begin{aligned}
     E^2(z)\rightarrow (1+z)^3 \left[\Omega_{b0}+\Omega_{r0}+z\Omega_{r0}-(\Omega_{b0}+\Omega_{r0}-1) \left(1+z\right)^{-3 \left[\frac{\Omega_{c0}}{\Omega_{b0}+\Omega_{r0}-1}+1\right]}\right].
   \end{aligned}
\end{equation}
\\
\subsubsection{MODEL 5. $Q_5= 3H \gamma \frac{\rho_{c}^2}{\rho}$}

The second non-linear term to be studied has the following dependence on the energy densities
\begin{equation}
    Q= 3H \gamma \frac{\rho_{c}^2}{\rho},\label{Q5}
\end{equation}
where $\rho=\rho_{c}+\rho_{de}$. In this case, the analytical solutions for the ratio $\tilde{r}$ and 
total energy density $\rho$ are
\begin{align}
  \tilde{r}&=\tilde{r}_0   \frac{w_{de}}{(w_{de}+\gamma \tilde{r}_0)a^{-3w_{de}}-\gamma \tilde{r}_0},\\
\rho&=\rho_0a^{-3\left(1-\frac{\gamma w_{de}}{w_{de}-\gamma}\right)}\left[\frac{(w_{de}+\gamma \tilde{r}_0)   a^{-3w_{de}}+\tilde{r}_0 (w_{de}-\gamma)}{w_{de}(1+\tilde{r}_0)}\right],  
\end{align}
then, the expression for the dimensionless Hubble rate is 
\begin{equation}\label{322}
    E^2(z)=\frac{\left(\frac{1}{1+z}\right)^{-3+w_{de} \left(-3+\frac{3 \gamma }{w_{de}-\gamma}\right)} \left(w_{de}\Omega_{de,0}-\gamma \Omega_{c0} \left(\left(\frac{1}{1+z}\right)^{3w_{de}}-1\right)\right)}{w_{de}}
    +\Omega_{c0} \left(\frac{1}{1+z}\right)^{\frac{3 \gamma  w_{de}}{w_{de}-\gamma}-3} 
    +\Omega_{b0}(1+z)^3+\Omega_{r0}(1+z)^4.
\end{equation}
For $a \ll 1$ (the high-redshift limit), the ratio $\tilde{r}$ becomes a constant, $\tilde{r} \rightarrow 
\left | w_{de} \right |/\gamma$. In the opposite limit ($a \gg 1$), $\rho \propto a^{-3}$, as in the $\Lambda$CDM case. Similarly for MODEL 4 when $\gamma=0$ and $w_{de}= -1$, $\Lambda$CDM is recovered.\\
A special case corresponds for $w_{de} \rightarrow -1$  and $\gamma \rightarrow 1$. The expression for the dimensionless Hubble rate (\ref{322}) exhibits a convergent behavior to
{\small \begin{align}\nonumber
E^2(z) &\rightarrow (1+z)^3\times \left[\Omega_{b0}+\left(\frac{1}{1+z}\right)^{\frac{3}{2}}\Omega_{c0}+\Omega_{r0}(1+z)-\left(\frac{1}{1+z}\right)^{\frac{9}{2}}\left([2-(1 + z)^3]\Omega_{c0}+\Omega_{b0}+\Omega_{r0}-1\right) \right].
\end{align}}

%\\
\subsubsection{MODEL 6. $ Q_6 = 3 H \gamma \frac{\rho_{de}^2}{\rho}$ }

The third interaction term belonging to this category is
\begin{equation}
    Q=3H \gamma \frac{\rho_{de}^2}{\rho}\label{Q6},
\end{equation}
where $\rho=\rho_{c}+\rho_{de}$. For $w_{de} < 0$, i.e. $w_{de} =-\left | w_{de} \right |$  the solutions are
\begin{align}
   \tilde{r}&=\left(\tilde{r_0}- \frac{\gamma}{\left | w_{de} \right |}\right )a^{-3\left | w_{de} \right |}+ \frac{\gamma}{\left | w_{de} \right |},\\
   \rho&=\rho_0a^{-3\left(1-\frac{w_{de}^2}{\left | w_{de} \right |+\gamma}\right)}\left[\frac{\left | w_{de} \right |+\gamma +(\left | w_{de} \right |\tilde{r_0}-\gamma)  a^{-3\left | w_{de} \right |}}{\left | w_{de} \right |(1+\tilde{r_0})}\right]^\frac{\left | w_{de} \right |}{\left | w_{de} \right |+\gamma },
\end{align}
while the dimensionless Hubble rate yields
\begin{flalign}\label{323} 
E^2(z)&=(z+1)^3\times &&
\end{flalign} 
{\tiny \begin{align}\nonumber
&\left[\Omega_{b0}+\Omega_{r0}+z\Omega_{r0}
  +(\Omega_{c0}+\Omega_{de,0})\left|w_{de}\right| ^{-\frac{\left|w_{de}\right|}{\left|w_{de}\right|+\gamma}}
  \left(\frac{1}{1+z}\right)^{\frac{3 w_{de}^2}{\left| w_{de}\right| +\gamma }}
  \left(\frac{\Omega_{de,0} (\left| w_{de}\right| +\gamma)+\left(\frac{1}{1+z}\right)^{-3 \left| w_{de}\right| } (\Omega_{c0} \left|w_{de}\right| -\gamma  \Omega{de,0})}{\Omega_{c0}+\Omega_{de,0}}\right)^{\frac{\left|w_{de}\right| }{\left|w_{de}\right| +\gamma }}\right].
\end{align}}

In this case, the expression (\ref{323}) is always defined for any values of $\gamma$ and $w_{de}$. Additionally, the ratio $\tilde{r}$ scales as $a^{-3\left|w_{de}\right|}$ for $a \ll 1$. For $w_{de} = -1$, this coincides with the scaling of  its $\Lambda$CDM counterpart. In the far-future limit, however, we obtain $\tilde{r} \Rightarrow \gamma/\left|w_{de}\right| $, $(a \gg 1)$, i.e., the energy-density ratio remains finite, while it tends to zero in the $\Lambda$CDM model.
The energy density scales as $a^{-3}$ for $a\ll 1$, i.e., we recover an early matter-dominated period. In the limit $a \gg 1$ one has 

\begin{equation}
    \rho \propto a^{-3\left(1-\frac{w_{de}^2}{\left| w_{de}\right| +\gamma}\right)}.
\end{equation}
From this last Eq., we can see that this generally does not correspond to a dS phase.

\subsection{Models with a change of direction of energy transfer}

In Ref.\cite{Cai:2009ht}, the authors investigated the interaction regardless the explicit form of $Q$, by using the most current observational data available at that time. The authors divided the whole redshift range $z$ into a few and concluded that the interaction term $\delta(z) = Q/(3H)$ was a constant in each redshift interval. They realized that $\delta(z)$ is likely to cross the line where the interaction does not occur ($\delta = 0$), therefore, the sign of the interaction $Q$ must change in the approximate redshift range of $0.45 \lesssim z \lesssim 0.9$.\\
This result poses a problem since most of the interactions studied in the literature are of linear type as in MODELS 1 and 2, which are always positive or negative and, therefore, cannot give the possibility of changing their signs. Taking into account the work \cite{Cai:2009ht}, the author proposed a new type of interaction in \cite{Wei:2010fz} and \cite{hao2011cosmological}, where the deceleration parameter $q$ was included in the coupling $Q$, thus allowing that the interaction $Q$ changes sign when the Universe goes from deceleration ($q > 0$) to acceleration ($q < 0$).
\\
In this Section, we will investigate MODELS 7, 8, and 9, previously studied in \cite{hao2011cosmological}, and later by \cite{bolotin2015cosmological,Arevalo:2022sne}. In these cases, $Q$ is proportional to the deceleration parameter, which produces a change of direction of energy transfer. For simplicity, in these works, the authors set $w_{de}=-1$.
\\
\subsubsection{MODEL 7. $Q_7= q(\alpha \dot{\rho}_{c} + 3 \beta H\rho_{c})$}
The first interaction within this category is a linear combination of energy density of DM and its time derivative
\begin{equation}
    Q= q(\alpha \dot{\rho}_{c} + 3 \beta H\rho_{c}).\label{Q7}
\end{equation}
For this kind of coupling and using Eq.(\ref{q2}) and (\ref{f2}), a transcendental differential equation for $H(a)$ is obtained
\begin{equation}
    a\frac{d^2H}{da^2}+\frac{a}{H}\left(\frac{dH}{da}\right)^2+\frac{dH}{da}=\frac{\beta q-1}{1-\alpha q}\cdot 3\frac{dH}{da}.
\end{equation}
which cannot be solved analytically for $\alpha \neq 0$. Nevertheless, in the particular case of $\alpha=0$, the analytic solution for the dimensionless Hubble rate is found to be
\begin{equation}\label{331}
E^2(z)=\left[1-\frac{2+3\beta}{2(1+\beta)}\Omega_{c0}\left[1-(1+z)^{3 (1+\beta)}\right]\right]^{\frac{2}{2+3\beta}}.
\end{equation}
When $\beta=0$, it yields
\begin{equation}
E^2(z)=\Omega_{c0}(1+z)^3+ (1-\Omega_{c0}),
\end{equation}
corresponding to $\Lambda$CDM, where the contribution coming from baryons and radiation components is neglected.\\
We note that if $\beta \rightarrow -1$ or $\beta \rightarrow -\frac{2}{3}$, the expression (\ref{331}) becomes undefined.\\
\subsubsection{MODEL 8. $Q_8= q(\alpha \dot{\rho}_{tot} + 3 \beta H\rho_{tot})$}
In this second case, the coupling 
involves the sum of DM and DE densities, i.e $\rho=\rho_c+\rho_{de}$ and its time derivative 
\begin{equation}
    Q= q(\alpha \dot{\rho} + 3 \beta H\rho).
\end{equation}
Similarly to the previous case, we arrive at a transcendental differential equation for $H(a)$
\begin{equation}
    a\frac{d^2H}{da^2}+\frac{a}{H}\left(\frac{dH}{da}\right)^2+(4+3\alpha q)\frac{dH}{da}+\frac{9\beta qH}{2a}=0,
\end{equation}
for $\alpha=0$, the interaction reduces to $Q= 3 \beta qH\rho_{tot}$, and the corresponding solution for the dimensionless Hubble rate yields 
\begin{equation}
E(z)=(1+z)^{3(2-3\beta +r_1)/8}\cdot\sqrt{\frac{C_{21}+(1+z)^{-3r_1/2}}{C_{21}+1}},
\end{equation}
where $r_1 \equiv \sqrt{4+\beta  (4+9 \beta)}$ and 
\begin{equation}
    C_{12}=-1+\frac{2r_1}{2-3 \beta-4\Omega_{c0}+r_1}.
\end{equation}
If $\beta=0$, $E(z)$ reduces to
\begin{equation}
E^2(z)=\Omega_{c0}(1+z)^3+ (1-\Omega_{c0}),
\end{equation}
which corresponds to $\Lambda$CDM.

\subsubsection{MODEL 9. $Q_9= q(\alpha \dot{\rho}_{de} + 3 \beta H\rho_{de})$}
\bigskip
Now, we are going to present a model where $Q$ is a linear combination of DE energy density and its time derivative
\begin{equation}
    Q= q(\alpha \dot{\rho}_{de} + 3 \beta H\rho_{de}).
\end{equation}
In the same way as two previous cases, we arrive at a transcendental differential equation for $H(a)$
\begin{equation}
    a\frac{d^2H}{da^2}+\frac{a}{H}\left(\frac{dH}{da}\right)^2+\left(4+\frac{3\beta q}{1+\alpha q}\right)\frac{dH}{da}+\frac{9\beta qH}{2a(1+\alpha q)}=0.
\end{equation}
We solve the equation for $\alpha=0$ and obtain the following dimensionless Hubble rate  
\begin{equation}\label{333}
E(z)=(1+z)^{3(2-5\beta +r_2)/[4(2-3\beta)]}\cdot\left[{\frac{(1+z)^{-3r_2/2}+C_{31}}{1+C_{31}}}\right]^{1/(2-3\beta)},
\end{equation}
where $r_2 \equiv \left | 2-\beta \right |$ and 
\begin{equation}
    C_{31}=-1+\frac{2r_2}{2- 5\beta+r_2+2\Omega_{c0}(3\beta-2)}.
\end{equation}
If we set $\beta=0$  
\begin{equation}
E^2(z)=\Omega_{c0}(1+z)^3+ (1-\Omega_{c0}),
\end{equation}
$\Lambda$CDM model is recovered.\\
If $\beta \rightarrow \frac{2}{3}$, the expression (\ref{333}) becomes
\begin{equation}
E^2(z)\rightarrow (1+z)^3 \exp{\left[{\frac{3z(z+2)(\Omega_{c0}-1)}{2(1+z)^2}}\right]}.   
\end{equation}

\subsection{Interaction involving derivatives of the energy densities}

\subsubsection{MODEL 10. $Q_{10}=3\alpha H (\rho_{de}'+\rho_{c}')$}

In Ref.\cite{Grandon:2018hom}, the authors studied an interaction where $Q$ is constructed to be a lineal combination of DE and DM  energy densities but involving derivatives (derivatives with respect to ln($a^3$), where $a$ is the scale factor) as
\begin{equation}
    Q=3\alpha H (\rho_{de}'+\rho_{c}'),
\end{equation}
where $ '\equiv \frac{d}{d\mathrm{ln}(a/a_0)^3}=\frac{d}{3Hdt}$. This model was previously studied in \cite{Chimento:2009hj} and later in \cite{Sharov:2017iue}. Assuming that $w_{de}$ is constant, in Ref. \cite{Grandon:2018hom} the authors provide an analytical solution for the total energy density $\rho=\rho_{de}+\rho_{c}$ and hence the analytical expression for $E^2(z)$. Firstly, the solution for $\rho$ is given by
\begin{equation}
    \rho(a)=C_1a^{3\beta^+}+C_2a^{3\beta^-},
\end{equation}
where
\begin{equation}\label{bet}
\beta^{\pm}=\frac{-2-(1-\alpha)w_{de} \pm \sqrt{(1-\alpha)^2w_{de}^2-4\alpha w_{de}}}{2},
\end{equation}
and 
\begin{equation}
F_\pm=\frac{\Omega_{de}(1+w_{de}+\beta^{\pm})+\Omega_c(1+\beta^{\pm})}{\beta^{-}-\beta^{+}}.
\end{equation}
We assume that $(1-\alpha)^2w_{de}^2-4\alpha w_{de} > 0$ for $\beta^{\pm}$ to be real and distinct. If $w_{de}=-1$ and $\alpha=-1$ implies $\beta^{+}=\beta^{-}$, because $(1-\alpha)^2w_{de}^2-4\alpha w_{de} = 0$. In that case, $F_\pm$ is undefined.
Thus, the dimensionless Hubble rate can be written as 
\begin{equation}
E^2(z)=\Omega_b(1+z)^3+\Omega_r(1+z)^4+(1+z)^{-3\beta^{+}}F_{-}-(1+z)^{-3\beta^{-}}F_{+}.
\end{equation}
It is easy to verify that if $\alpha=0$ there is no interaction. We obtain from Eq. (\ref{bet}), $\beta^{+}=-1$, and $\beta^{-}=-(1+w_{de})$, therefore $F_{+}=-\Omega_x$ and $F_{-}=\Omega_c$, therefore, the model reduces to $w$CDM.

\subsection{Parameterized interactions}

In the following Subsection, we present some interactions which are derived from a real function $f(\tilde{r})$ of the coincidence parameter $\tilde{r}$ is introduced, which is the ratio between the energy densities of DM and DE, and it is given by Eq.(\ref{rtilde}). The main motivation for studying the interactions parametrically is to solve the cosmic coincidence problem. By solving the differential equations one has an analytical solution for the ratio, allowing us to see the evolution of the ratio as a function of the scale factor or redshift. Some of the interactions examined in the preceding models can be obtained by appropriately selecting the parametric function $f(\tilde{r})$. \\
In Ref.\cite{cosmoconstrain}, five models were investigated by the authors. In this reference is set $w_{de}=-1$ and the sign of $Q$ is changed, hence, the equations (\ref{q1}) and (\ref{q2}) have the following form
\begin{equation}\label{q11}
\dot{\rho}_{c}+3H\rho_{c}= -Q,
\end{equation}
\begin{equation}\label{q12}
\dot{\rho}_{de}=Q,
\end{equation}
where $Q$ is written as $Q=3H\gamma R(\rho_c, \rho_{de})$, whit $\gamma$ being a dimensionless constant and $R$ is a real function which has 
units of energy density.
\\
By taking the time derivative of the ratio $\tilde{r}$ between the energy density of DM and DE as Eq. (\ref{rtilde}), we obtain the expression
\begin{equation}\label{ratio}
\dot{\tilde{r}}=\tilde{r} \left(\frac{\dot{\rho_c}}{\rho_c}-\frac{\dot{\rho}_{de}}{\rho_{de}}\right).
\end{equation}
Using the equations (\ref{q11}), (\ref{q12}) and (\ref{ratio}) together, we get
\begin{equation}\label{diff}
\dot{\tilde{r}}+3H\tilde{r}[\gamma f(\tilde{r})+1]=0,
\end{equation}
where 
\begin{equation}
f(\tilde{r})=R \frac{\rho_c+\rho_{de}}{\rho_c\rho_{de}}.
\end{equation}
We are interested in the analytical solution for (\ref{diff}), and we can obtain the expression for $\rho_c=\rho_c(a)$, $\rho_{de}=\rho_{de}(a)$ when $\tilde{r}=\tilde{r}(a)$.

The following models correspond to different choices of $f(\tilde{r})$.

\subsubsection{MODEL 11. $f(\tilde{r})=1$}

The first parametric form is chosen to be
\begin{equation}
    f(\tilde{r})=1,
\end{equation}
which is equivalent to the interaction of MODEL 4, given by Eq.(\ref{Q4})
\begin{equation}
   Q= 3H \gamma \frac{\rho_{c}\rho_{de}}{\rho_{c}+\rho_{de}}. 
\end{equation}
Besides, the coincidence parameter has the following solution
\begin{equation}
\tilde{r}(a)=\tilde{r}_0a^{-3(\gamma+1)},
\end{equation}
while the energy density of DM yields
\begin{equation}
    \rho_c=\rho_{c0}a^{-3}\left(\frac{\Omega_{c0}+\Omega_{de,0}a^{3(\gamma+1)}}{\Omega_{c0}+\Omega_{de,0}}\right)^{-\frac{\gamma }{\gamma +1}}.
\end{equation}
By using the relation $a=(1+z)^{-1}$, we have the expression for the dimensionless rate Hubble as follows
\begin{equation}\label{f(r)=1}
 E^2(z)=(\Omega_{c0}+\Omega_{de,0})\left(\frac{\Omega_{c0}+\Omega_{de,0}(1+z)^{-3(\gamma+1)}}{\Omega_{c0}+\Omega_{de,0}}\right)^{\frac{1}{1+\gamma}}(1+z)^{3}+\Omega_{b0}(1+z)^{3}+\Omega_{r0}(1+z)^{4}.
\end{equation}
It is important to mention that when we analyze the models numerically in next Section, the sign of $\gamma$ is the opposite of that in the models with equivalent interactions since the sign for $Q$ has been chosen with the opposite sign.\\
In the same way as we did in MODEL 4, we investigate the case where $w=\gamma=-1$. If $ \gamma \rightarrow -1$. In such a case, the dimensionless Hubble rate (\ref{f(r)=1}) takes the form
\begin{equation}
\begin{aligned}
     E^2(z)\rightarrow (1+z)^3 \left[\Omega_{b0}+\Omega_{r0}+z\Omega_{r0}-(\Omega_{b0}+\Omega_{r0}-1) \left(1+z\right)^{-3 \left[\frac{\Omega_{c0}}{\Omega_{b0}+\Omega_{r0}-1}+1\right]}\right],
   \end{aligned}
\end{equation}
which is equivalent to the equation (\ref{4red}). 

\subsubsection{MODEL 12. $ f(\tilde{r})=\frac{1}{\tilde{r}}$}

By choosing the following explicit form for $f(\tilde{r})$
\begin{equation}
    f(\tilde{r})=\frac{1}{\tilde{r}},
\end{equation}
results in a coupling $Q$ which is equivalent to the interaction of MODEL 6 given by Eq.(\ref{Q6})
\begin{equation}
   Q= 3H \gamma \frac{\rho_{de}^2}{\rho_{c}+\rho_{de}}, 
\end{equation}
and the solution for $\tilde{r}(a)$ is
\begin{equation}
\tilde{r}(a)=\tilde{r}_0a^{-3}-\gamma(1-a^{-3}).
\end{equation}
Besides, the analytical solution for the DE density is given as
\begin{equation}
   \rho_{de}=\rho_{de,0}a^{-\frac{3\gamma }{\gamma -1}} \left(\frac{(1-\gamma )\Omega_{de,0}+(\Omega_{c0}+\gamma\Omega_{de,0})a^{-3}}{\Omega_{c0}+\Omega_{de,0}}\right)^{-\frac{\gamma }{\gamma -1}},
\end{equation}
while the dimensionless Hubble rate is expressed as
\begin{equation}
\begin{aligned}
   E^2(z)= \Omega_{de,0} \left(\frac{1}{z+1}\right)^{-\frac{3 \gamma }{\gamma -1}} \left(\frac{(1-\gamma ) \Omega_{de,0}+(z+1)^3(\gamma\Omega_{de,0}+\Omega_{c0})}{1-\Omega_{b0}-\Omega_{r0}}\right)^{-\frac{\gamma }{\gamma -1}} \\    +\left(\frac{1}{z+1}\right)^{-\frac{3 \gamma }{\gamma -1}} \left(\gamma  \Omega_{de,0} z (z (z+3)+3)  
   +\Omega_{c0}(z+1)^3\right)
   \times \left(-\frac{(\gamma -1) \Omega_{de,0}+(z+1)^3 (\Omega_{c0}-\gamma  \Omega_{de,0})}{\Omega_{b0}+\Omega_{r0}-1}\right)^{-\frac{\gamma }{\gamma -1}} \\ 
  + \Omega_{b0} (z+1)^3+\Omega_{r0} (z+1)^4.
   \end{aligned}
\end{equation}

\subsubsection{MODEL 13. $f(\tilde{r})=\tilde{r}$}

An equivalent interaction as MODEL 5, given by Eq.(\ref{Q5})
\begin{equation}
   Q= 3H \gamma \frac{\rho_{c}^2}{\rho_{c}+\rho_{de}} 
\end{equation}
can be obtained if $f(\tilde{r})$ depends linearly on the coincidence parameter, i.e.
\begin{equation}
    f(\tilde{r})=\tilde{r}.
\end{equation}
For this parametrization, the coincidence parameter has the solution
\begin{equation}
\tilde{r}(a)=\tilde{r}_0\frac{a^{-3}}{1+\tilde{r}_0\gamma-\tilde{r}_0\gamma a^{-3}},
\end{equation}
and the DM energy density results in
\begin{equation}
  \rho_c=\rho_{c0}a^{-3}\left[\frac{\gamma\Omega_{c0}+(1-\gamma)\Omega_{c0}a^{-3}+\Omega_{de,0}}{\Omega_{c0}+\Omega_{de,0}}\right]^{-\frac{\gamma }{\gamma -1}}.
\end{equation}
In this way, the dimensionless Hubble rate is written as
\begin{equation}
    \begin{aligned}
E^2(z)= \Omega_{c0}(z+1)^3 \left(\frac{\gamma \Omega_{c0}+\Omega_{de,0}+(1-\gamma ) \Omega_{c0} (z+1)^3}{1-\Omega_{b0}-\Omega_{r0}}\right)^{-\frac{\gamma }{\gamma -1}}\\
+\left(\gamma  \Omega_{c0}+\Omega_{de,0}-\gamma  \Omega_{c0}(z+1)^3\right)
\times \left(\frac{\gamma  \Omega_{de,0}+(1-\gamma ) \Omega_{c0}(z+1)^3}{1-\Omega_{b0}-\Omega_{r0}}\right)^{-\frac{\gamma }{\gamma -1}}
+ \Omega_{b0}(z+1)^3+\Omega_{r0}(z+1)^4.
    \end{aligned}
\end{equation}

\subsubsection{MODEL 14. $f(\tilde{r})=1+\frac{1}{\tilde{r}}$}
For a parameterization of the type
\begin{equation}
    f(\tilde{r})=1+\frac{1}{\tilde{r}},
\end{equation}
the corresponding rate $Q$ becomes equivalent to MODEL 2, which is given by Eq.(\ref{Q2})
\begin{equation}
   Q= 3H \gamma \rho_{de},
\end{equation}
and the coincidence parameter may be written as follows
\begin{equation}
\tilde{r}(a)=a^{-3 \gamma}\frac{ \left(\tilde{r}_0a^{-3}+\gamma a^{-3}- \gamma a^{3\gamma}+\tilde{r}_0\gamma a^{-3}\right)}{1+\gamma}.
\end{equation}
Therefore, the dimensionless Hubble rate becomes
\begin{equation}
    \begin{aligned}
    E^2(z)=\frac{\gamma \Omega_{de,0} \left(-\left(\frac{1}{z+1}\right)^{3 \gamma }\right) +\gamma \Omega_{c0} (z+1)^3+\Omega_{c0}(z+1)^3
    +\Omega_{de,0}\gamma (z+1)^3}{\gamma +1}\\
        +\Omega_{de,0} \left(\frac{1}{z+1}\right)^{3 \gamma }+\Omega_{b0}(z+1)^3+\Omega_{r0}(z+1)^4.
    \end{aligned}
\end{equation}

\subsubsection{MODEL 15. $f(\tilde{r})=1+\tilde{r}$}
The interaction of the MODEL 1 
\begin{equation}
   Q= 3H \gamma \rho_{c},
\end{equation}
can be reproduced if $f(\tilde{r})$ has the following explicit dependence on $\tilde{r}$
\begin{equation}
    f(\tilde{r})=1+\tilde{r}.
\end{equation}
For this last parametrization, the solution for the ratio $\tilde{r}$ is found to be
\begin{equation}
\tilde{r}(a)= -\frac{1+\gamma}{\gamma-a^{3(1+\gamma)}\left(\frac{1+\gamma+\tilde{r}_0 \gamma}{\tilde{r}_0} \right)},
\end{equation}
and accordingly
\begin{equation}
  \rho_c=\rho_{c0}a^{-3(1+\gamma)}.
\end{equation}
In this model, the dimensionless Hubble rate yields
\begin{equation}
   E^2(z)= \frac{\Omega_{de,0}+\gamma \left(\Omega_{c0}+\Omega_{de,0}-\Omega_{c0} \left(\frac{1}{z+1}\right)^{-3 (\gamma +1)}\right)}{\gamma +1}+\Omega_{c0} \left(\frac{1}{z+1}\right)^{-3 (\gamma +1)}
   +\Omega_{b0}(z+1)^3+\Omega_{r0} (z+1)^4.
  \end{equation}
We observe that $E^2(z)$ reduces to $\Lambda$CDM when $\gamma=0$.

\subsection{Self-interaction between DM}

In Ref.\cite{Cardenas:2020nny}, the authors instead of considering a DE component, they argue that a self-interacting DM produces the accelerated expansion. They investigate background cosmic evolution by assuming a direct interaction between two DM particle types. One species is denoted by a $m$ subscript, while the other one is denoted by a $x$ subscript.
It is important to point out that baryons and radiation are not considered in this work as self-interacting models for small redshift will be studied.
\\
By defining a Langrangian density, conservation equations for each type of DM particles are derived, which are similar to the equations (\ref{q1}) and (\ref{q2}), thus
\begin{equation}\label{qm}
\dot{\rho}_{m}+3H(\rho_{m}+p_m)=Q_m,
\end{equation}
\begin{equation}\label{qx}
\dot{\rho}_{x}+3H(\rho_{x}+p_x)=Q_x.
\end{equation}
In that work, two specific models are studied: a symmetric model and an asymmetric model with respect to the $Q_m$ and $Q_x$ interaction rates.

\subsubsection{MODEL 16. {\small Symmetric model. $Q_m=-3H\alpha\rho_m, ~~~ Q_x=-3H\beta\rho_x$}}

In the first model, a simple linear interaction between the both DM components is proposed, and it is similar to those investigated between DE and DM interacting models, with the benefit of knowing the mathematical behavior through the analytical solution of the density equations. As a result, the interaction rates are defined as follows
\begin{equation}
Q_m=-3H\alpha\rho_m, ~~~ Q_x=-3H\beta\rho_x,
\end{equation}
thus, solving Eqs. (\ref{qm}) and (\ref{qx}), and assuming $p_m=p_x=0$, it was obtained the following dimensionless Hubble rate 
\begin{equation}
E^2(z)=1+\frac{\Omega_m}{\alpha +1}\left((z+1)^{3(\alpha +1)}-1\right)+\frac{\Omega_x}{\beta+1}\left((z+1)^{3(\beta +1)}-1\right),
\end{equation}
where $\Omega_i=\rho_i^0/3H_0^2$. As it can be seen, the equation for $E^2(z)$ is symmetrical for both types of DM components. It is crucial to note that when $\alpha, \beta \ll 1$ (but non null) implies that $E^2(z)$ reduces to 
\begin{equation}
E^2(z) \simeq 1-\Omega_m-\Omega_x+(\Omega_m+\Omega_x)(z+1)^{3}.
\end{equation}
This indicates that having a small (but non-zero) interaction between two dust-like components, the model we obtain is very similar to a model with no interaction containing dust and a cosmological constant.

\subsubsection{MODEL 17. Asymmetric model. $Q_m=-3H\alpha\rho_m, Q_x=-3H\beta(\rho_m+\rho_x)$}

Linear interaction functions may be defined in an asymmetric way, such that
\begin{equation}
Q_m=-3H\alpha\rho_m, \hspace{1cm} Q_x=-3H\beta(\rho_m+\rho_x),
\end{equation}
which implies that the solution for the corresponding Hubble rate yields
\begin{equation}
E^2(z)=1+\Omega_m [ [(\alpha +\beta)-1] \left(1-(z+1)^3\right) -3\beta(1+z)^3 \log(1+z)] +\Omega_x \left[(\alpha +\beta) -1\right ] \left(1-(z+1)^3 \right).
\end{equation}
In the limit $z \rightarrow -1$ for the expression $E^2(z)$, we get 
\begin{equation}
E^2(z) \rightarrow 1-(\Omega_m +\Omega_x)-3\beta\Omega_m \left[(z+1)^3\log(1+z)\right],
\end{equation}
clearly the term between the square brackets vanishes when $z \rightarrow -1$. It is concluded that, regardless of the value of $\beta$, in the limit when $z \rightarrow -1$, $\Lambda$CDM is recovered.

\section{Statefinder analysis of interacting DE models}

Previously, we have shown how to obtain the dimensionless Hubble rate $E(z)$ for the several interacting models. Now, in the present Section we will compute the Statefinder parameters $q(z), r(z)$, and $s(z)$ for all the interacting DE models presented in previous section. We present the equations for $q(z), r(z)$, and $s(z)$ for each model and will not be shown due to that, generally, they are quite large. We will also obtain the present and asymptotic values of them as it is shown in the Table (\ref{statevalues}).
\\
We will analyze the models for three important epochs, namely, early epoch ($z \gg 1$), present epoch ($z=0$), and future epoch ($z \rightarrow -1$). As we focus on the evolution for late times, particularly in the transition from the matter-dominated era to the present epoch, we neglect the radiation component in the numerical calculation of $E(z)$ and the Statefinder $\{q, r, s\}$. 
In the subsequent figures, we plot the behavior of the evolutionary trajectories in the $q-r$ and $s-r$ planes, in order to compare the performance of the $q-r$, $s-r$ parametric graphs for each model with $\Lambda$CDM. The plots corresponding to (\ref{figurL:4}) and (\ref{figurL:5}) will overlay with the curves evaluated with the fitted values of the normalized densities and constants for the respective model. With the values of the Statefinder parameters that we have calculated, we will be able to establish the deviations of this model from $\Lambda$CDM \cite{Sahni:DMDE}. On this subject, we will say that an interacting model is \textit{compatible} with $\Lambda$CDM when for the matter-dominated era, the present time, and in the future, the Statefinder parameters are exactly or very close to those predicted by $\Lambda$CDM. When this is only satisfied for one o two epochs, we will say that it is \textit{partially compatible} and if the Statefinder parameters deviate significantly from those predicted by $\Lambda$CDM at all epochs or if there exists a large deviation in the matter-dominated era, we will say that the model is \textit{incompatible}.
In order to draw the plots of $H(z)$ with error bars, we use the values from Ref. \cite{meng2015utility} with $0\leq z \leq 2.36$.

\subsection{Linear models}

\subsubsection{MODEL 1. $Q_1=3\gamma H \rho_c$}

\begin{figure}[htp]
  \centering
  \label{figur}
\subfloat[]{\label{figurc:1}\includegraphics[width=80mm]{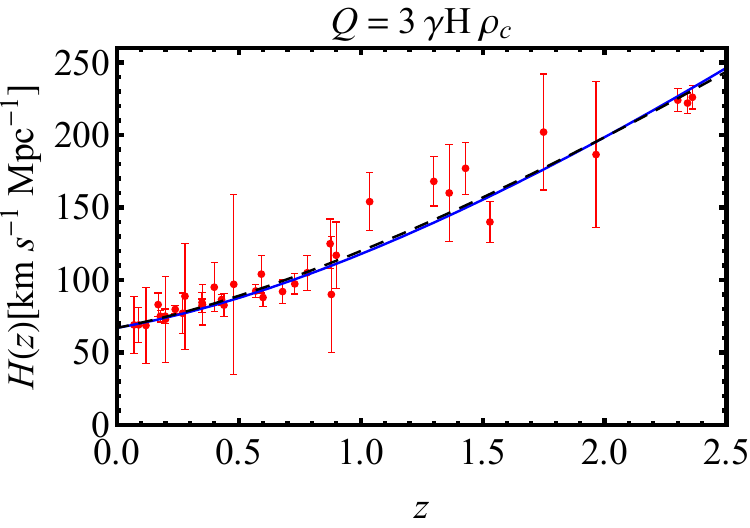}}
  \\
\subfloat[]{\label{figurc:2}\includegraphics[width=80mm]{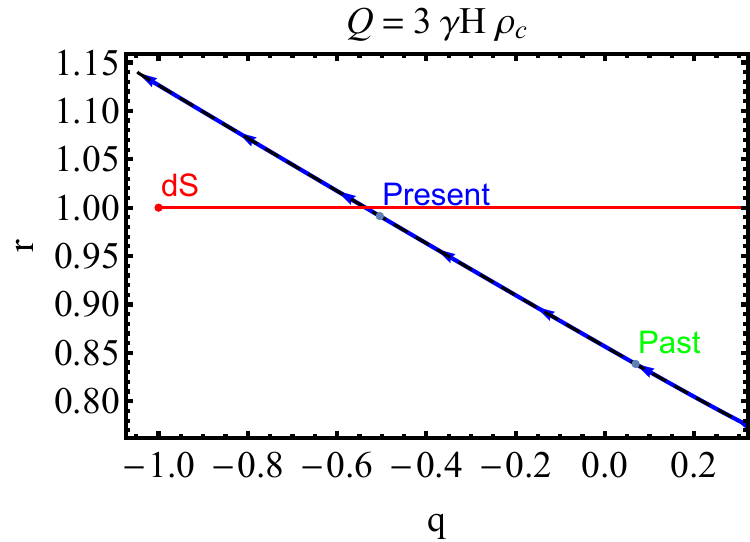}}
\subfloat[]{\label{figurc:3}\includegraphics[width=80mm]{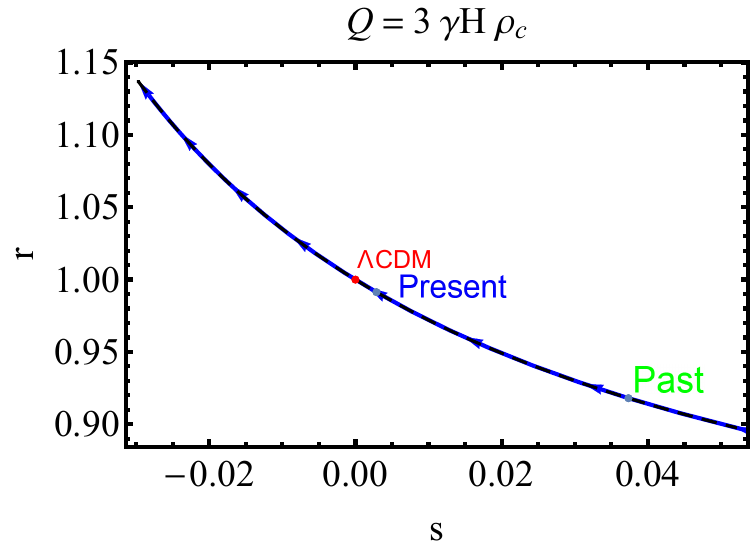}}
\caption{{\small In Figure (a) it is shown the Hubble rate $H(z)$ against $z$, for the model with the coupling function $Q_1 = 3\gamma H \rho_c$, representing the interaction between DE and DM (dashed line), as well as, the Hubble rate for $\Lambda$CDM model (solid line), along with the data from Ref. \cite{meng2015utility}. We have used  $H_0=66.9 {~}\mathrm{km~s^{-1}~Mpc^{-1}}$, from \cite{Grandon:2018hom}. For this model, Figures (b) and (c) show the $q - r$ and $s - r$, planes, respectively. The arrows on the curves show the direction of evolution.}}
\end{figure}

We extract the values of the normalized densities from Ref. \cite{Grandon:2018hom} in order to evaluate this model numerically. We have employed the values $h=0.669$, $\Omega_{c}=0.301$, $\Omega_{b}=0.049$, $\gamma=0.071$, $w_{de}=-1.03$, corresponding to SNIa+$H(z)$+BAO+$f_{gas}$+CMB, Table II, column C.\\
As we can see in the graph of Fig.(\ref{figurc:1}), the number of measurements for high redshift is decreasing, also we can see that for redshift higher than $0.5$ the uncertainty of the measurements increases. In general the $H(z)$ curves obtained for each model fit accurately to the points with error bars. Of these, we will only show a few for each model category since they are quite similar.
\\
In Fig.(\ref{figurc:2}) it is shown the parametric curve of MODEL 1 in the $(q, r)$ plane. For early times the trajectory  starts in the lower right corner of the figure, in the region bounded by $0< q < 1$ and $0<r$. When $z \gg 1$, the parameters $\left\{ q,r,s \right\}  \rightarrow \left \{0.42,0.79, 0.93 \right \}$, hence, the values for $q$, $r$, and $s$ deviate significantly in the standard matter-dominated era from $\Lambda$CDM, since for $\Lambda$CDM, the values for $q, r$ and $s$ are given by $q=0.5$, $r=1$ and $s=0$, respectively. However, the value obtained for $s$ is close to $1$, which corresponds to SCDM \cite{Sahni:DMDE}, \cite{Alam:2003sc}.\\
We have evaluated the model in the past at $z=0.8$ as a benchmark. For the subsequent parametric plots of all models, we will follow the same procedure; we will use different redshift values associated to the past. We plot that point, in order to better observe the temporal evolution of the model.
\\
At the present time, $z =0$, we obtain $\left\{ q,r,s \right\}  = \left \{-0.50,0.99,0.003 \right \}$. 
It is evident that the curve has a linear behavior and evolves in the negative direction of $q$ and positive of $r$. We note that close to the present time, the curve crosses the red straight line $r=1$ corresponding to $\Lambda$CDM. The value of $q$ deviates significantly from the value of  $q_0\simeq-0.55$, associated to $\Lambda$CDM. The values of $r$ and $s$ are very close from the values of  $r_0=1$ and $s_0=0$, obtained for  $\Lambda$CDM.
\\
At late times, the curve evolves and moves away from the dS fixed point in the future ($q = -1; r = 1$), as we have calculated in the limit $z \rightarrow -1$, we have $\left\{ q,r,s \right\}  \rightarrow \left \{ -1.0, 1.1,-0.03 \right \}$. Thus, we conclude that the model does not converge to $\Lambda$CDM in the future.\\
In Fig. (\ref{figurc:3}) we see the $(s, r)$ plane and we note that the plot starts in the region bounded by $0<s< 1$ and $r > 0$. The trajectory is non-linear and proceeds in the negative direction of $s$ and positive of $r$. We corroborate that around the present time, it crosses the fixed point $(0,1)$ corresponding to $\Lambda$CDM. According to our analysis, we will say that this model is partially compatible with $\Lambda$CDM.

\subsubsection{MODEL 2. $Q_2=3\gamma H \rho_{de}$}

\begin{figure}[htp]
  \centering
  \label{figur}
\subfloat[]{\label{figur_DE:0}\includegraphics[width=80mm]{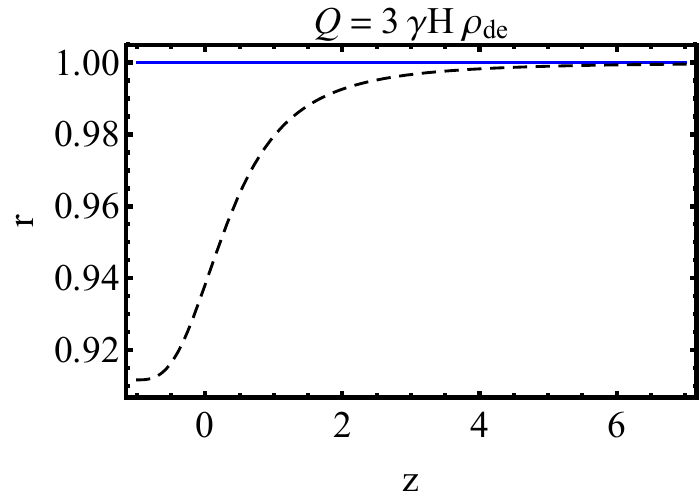}}
\subfloat[]{\label{figur_DE:1}\includegraphics[width=80mm]{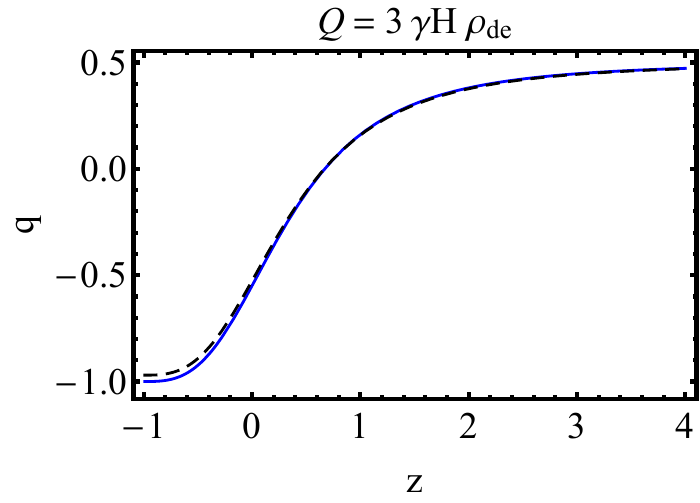}}
\\
  \subfloat[]{\label{figur_DE:2}\includegraphics[width=80mm]{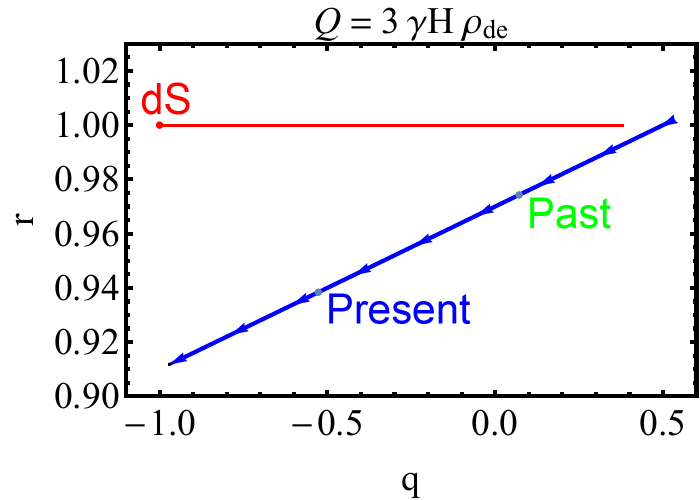}}
  \subfloat[]{\label{figur_DE:3}\includegraphics[width=80mm]{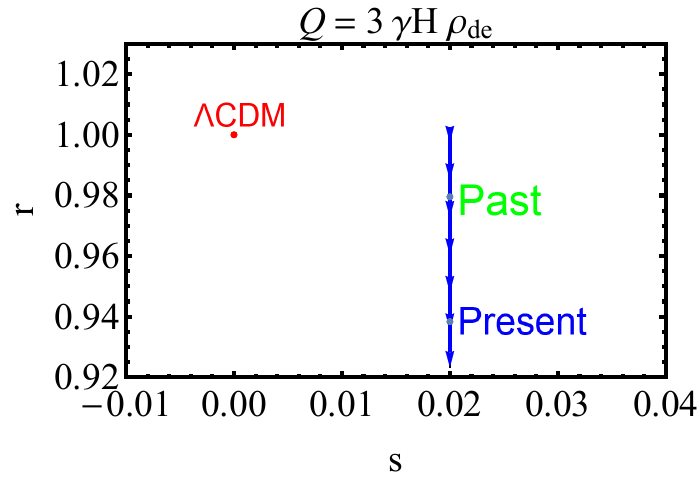}}
  
\caption{{\small In this figure we show the plots of the Statefinder parameters for the coupling function $Q_2 = 3\gamma H \rho_{de}$. Figs. (a) and (b) show the evolution of $r(z)$ and $q(z)$, respectively. We represent the interacting model DE-DM (black dashed line), which overlaps with $\Lambda$CDM (blue solid line).
For this model, the figures (c) and (d) show the $q - r$ and $s-r$ planes, respectively.}} 
\end{figure}
In order to evaluate this model numerically, we used the values of the normalized densities from Ref. \cite{Grandon:2018hom}. We will use SNIa+$H(z)$+BAO+$f_{gas}$+CMB, Table I, column C. $h=0.671$, $\Omega_{c}=0.300$, $\Omega_{b}=0.048$, $\gamma=0.07$, $w_{de}=-1.05$. 
\\
When $z \gg 1$, the parameters $\left\{q,r,s \right\}  \rightarrow \left \{0.5,1, 0.02 \right \}$, and we see that the value of $q$ and $r$ correspond to the standard matter-dominated era, and $s$ is very close to the value $s=0$ for $\Lambda$CDM. We observe in the Figs. (\ref{figur_DE:0}) and (\ref{figur_DE:1}) the asymptotic behaviour of $r(z)$ and $q(z)$ for high redshift.\\
For MODEL 2 in the $(q, r)$ plane, we show in Fig.(\ref{figur_DE:2}) that the trajectory is linear and starts in the upper right corner, in the region bounded by $0< q < 1$ and $0< r<1$.  Moreover, the curve evolves in the negative direction of $q$ and negative of $r$. At the present time, $z =0$, we obtain $\left\{ q,r,s \right\}  = \left \{-0.53,0.94, 0.02 \right \}$, which are very close values for the Statefinder parameters for $\Lambda$CDM.
When $z \rightarrow -1$ the parameters $\left\{ q,r,s \right\}  \rightarrow \left \{-0.97,0.91, 0.02 \right \}$. Thus, it is observed a significant deviation from the dS fixed point in the future. \\
In the $(s, r)$ plane, Fig.(\ref{figur_DE:3}) shows that the curve starts to the right side of the point corresponding to $\Lambda$CDM, in the region bounded by $0.01< s < 0.03$ and $0< r < 1$. The trajectory is linear in behaviour and proceeds in the negative direction of $r$, while holding fixed $s=0.02$. It is clearly observed that the behaviour of the curve in the $s-r$ plane is very similar to those of Quintessence models \cite{sahni2003statefinder}, \cite{Sahni:EXP}. It is observed that $s$ remains constant, while $r$ decreases asymptotically up to $r=0.91$. We corroborate the information provided by the $(q, r)$ plane; the trajectory of model deviates from $\Lambda$CDM.  
It is concluded that this model is partially compatible with $\Lambda$CDM.

\subsubsection{MODEL 3. $Q_3= 3\lambda_{c}H\rho_{c} +3\lambda_{de}H\rho_{de}$}

\begin{figure}[htp]
  \centering
  \label{figur}
  \subfloat[]{\label{figurCL:2}\includegraphics[width=80mm]{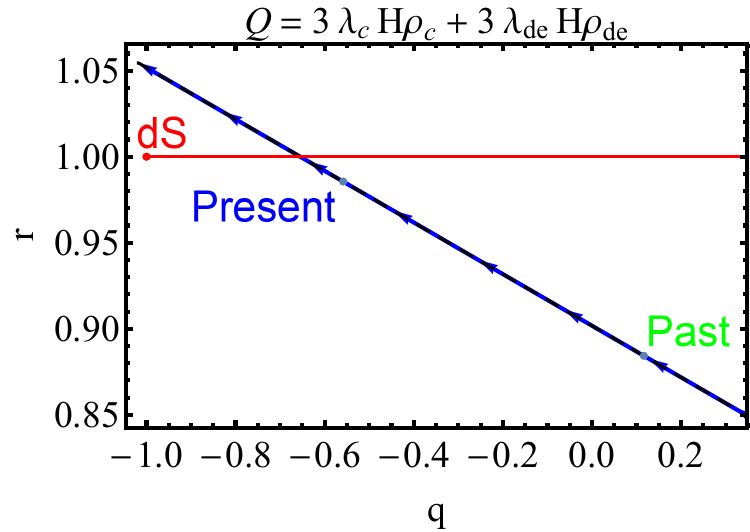}}
  \subfloat[]{\label{figurCL:3}\includegraphics[width=80mm]{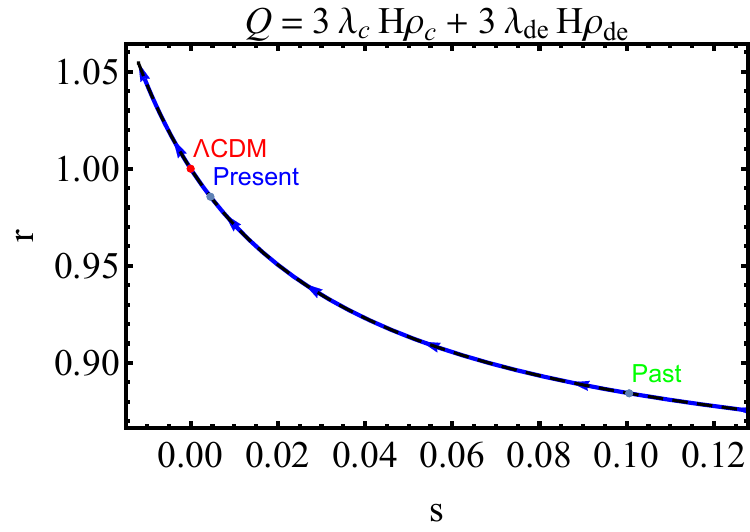}}
\caption{{\small \textbf{Left panel}: Parametric curve for coupling $Q_3= 3\lambda_{c}H\rho_{c} +3\lambda_{de}H\rho_{de}$ in $q - r$ plane. \textbf{Right panel}: Parametric curve in $s -r$ plane.}} \end{figure}

For this model we extract the values of the normalized densities and constants from Ref.\cite{Pan:2016ngu}. Table III. $\Omega_k = 0$, $\Omega_{m0} = 0.285$, $\lambda_c= 0.039$, $\lambda_{de} = -0.024$, $w_{de}= -0.987$.  
When we plot the curves on $r-s$ and $r-q$ planes, the baryons are considerate by separate (we choose for the model $ \Omega_{b0} \approx 0.04$) and the model has the same behaviour.  In \cite{Pan:2016ngu} the baryons are not considered separately in the tables. Also, radiation is not considered in their work but we consider it in the equations, however, we set it to zero when numerically evaluating the model.\\
For $z \gg 1$, the parameters $q, r$ and $s$ behaves as $\left\{ q,r,s \right\}  \rightarrow \left \{0.44,0.84,0.96 \right \}$. In this case $q$, and $r$ deviate significantly in the standard matter-dominated era for $\Lambda$CDM, but the value  $s$ is close to $1$, which corresponds to SCDM.\\
Fig.(\ref{figurCL:2}) shows the trajectory of MODEL 3 on the $(q, r)$ plane. This curve starts in the lower right corner of the figure, in the region bounded by $0< q < 1$ and $r > 0$. The trajectory has linear behaviour and proceeds in the negative direction of $q$ and positive of $r$. At the present time $z =0$, we obtain $\left\{ q,r,s \right\} = \left \{-0.56,0.99,0.005 \right \}$, 
which are very close values to, those expected for $\Lambda$CDM. \\
After the present time, the trajectory crosses the red straight line $r=1$ corresponding to $\Lambda$CDM, then it evolves very close to the dS fixed point in the future, which is consistent with the computation in the limit $z \rightarrow -1$. In this case the parameters $\left\{ q,r,s \right\}  \rightarrow \left \{-1,1.1, -0.012 \right \}$. 
Fig. (\ref{figurCL:3}) shows the $(s, r)$ plane, where the curve starts in the region bounded by $0< s < 1$ and $r > 0$. The trajectory is non-linear and evolves in the negative direction of $s$ and positive of $r$. We observe that after the current time, it crosses the fixed point $(0,1)$ corresponding to $\Lambda$CDM.\\ 
This model is partially compatible with $\Lambda$CDM, and it behaves similar to MODEL 1, which can be seen in the two parametric plots. This suggests that in the linear combination, the DM contribution is larger than the coming from DE.

\subsection{Non-linear interaction}
%\\
\subsubsection{MODEL 4. $Q_4= 3H \gamma \frac{\rho_{c}\rho_{de}}{\rho}$}

\begin{figure}[htp]
  \centering
  \label{figur}

  \subfloat[]{\label{figur321:1}\includegraphics[width=80mm]{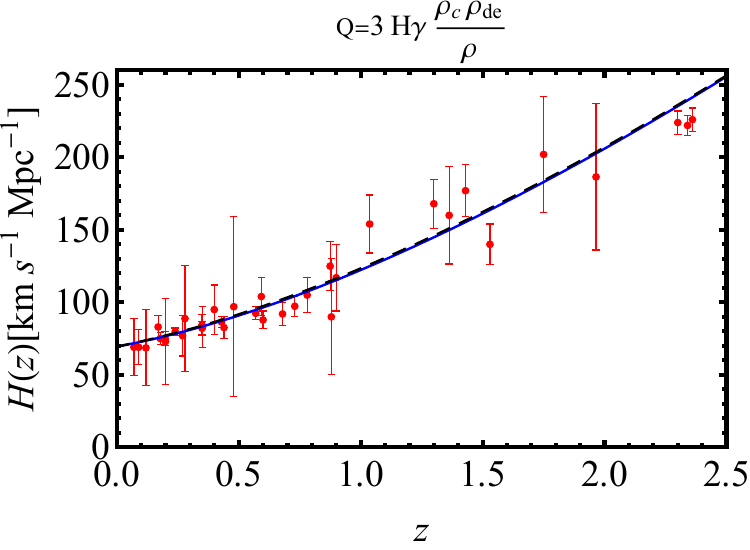}}
   \\
  \subfloat[]{\label{figur321:2}\includegraphics[width=80mm]{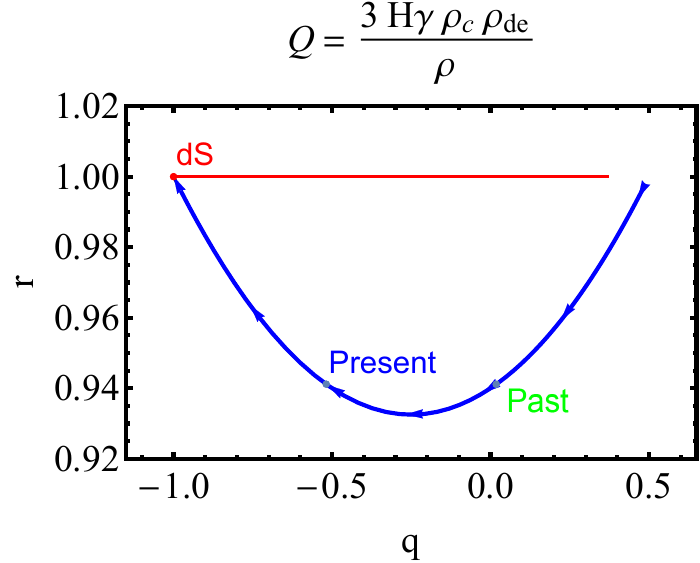}}
  \subfloat[]{\label{figur321:3}\includegraphics[width=80mm]{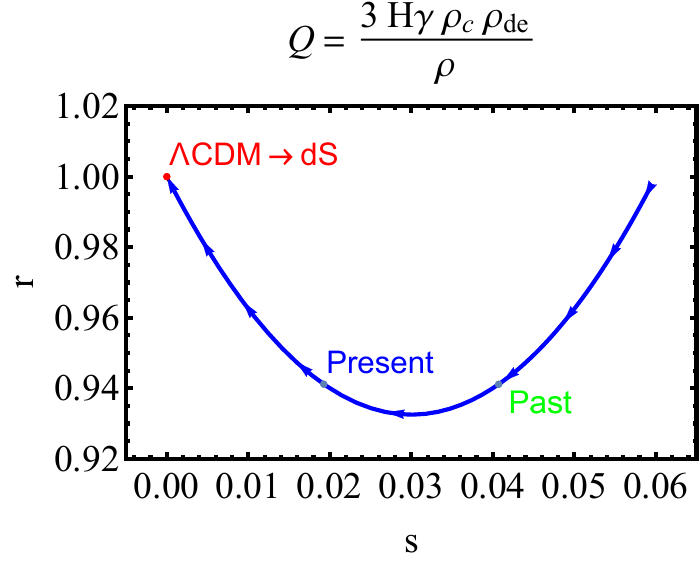}}
\caption{{\small Fig. (a) shows the Hubble rate $H(z)$ as a function of redshift $z$ for the model $Q_4= 3H \gamma \frac{\rho_{c}\rho_{de}}{\rho}$.  We have used  $H_0=69.44 {~}\mathrm{km~s^{-1}~Mpc^{-1}}$, from Ref. \cite{cosmoconstrain}. In Figures (b) and (c)  we show the parametric curves in the $q - r$ and $s -r$ planes, respectively.}}
\end{figure}
We observe in the $H(z)$ plot that for this model the curve moves away from the $H(z)$ values with the error bars for redshift greater than 2.\\
The values we will use to evaluate this model are obtained from the Ref. \cite{cosmoconstrain}, which we will cite below to investigate models where interactions are parameterized. For this model, we have used the values $\Omega_{m0}= 0.321$, $\gamma=-0.06$, $w_{de}=-1$, using SNe Ia + $H_0$ + CC + BAO, from Table 3 of \cite{cosmoconstrain}. However, in order to evaluate the non-linear MODELS 4, 5, and 6 numerically, we will use the value of $\gamma$ with the positive sign, since in \cite{cosmoconstrain}, the sign of $Q$ is changed when defining the conservation equations. Furthermore, the contribution coming from baryons is not considered separately in Table 3 of values in same work. When we evaluate our model numerically, we neglect baryons and assign all matter to DM, but to evaluate the model for $\Lambda$CDM we will separate baryons. When we consider baryons separately and evaluate our model numerically, the behaviour of the curves does not change significantly. \\
As it can be seen from Fig. (\ref{figur321:2}), that for MODEL 4 in the $(q, r)$ plane, the curve starts in the region bounded by $0< q < 1$ and $0< r<1$, in the upper right corner. For $z \gg 1$, the parameters $q, r$, and $s$ behaves as $\left\{ q,r,s \right\}  \rightarrow \left \{0.5,1,0.06\right \}$, where $q$ and $r$ are exactly as $\Lambda$CDM predicts, and $s$ is very close to zero, in this case we recover the standard matter-dominated era. The trajectory is non-linear in behaviour (like a concave-up parabola) and evolves in the negative direction of $q$ and at $r$ drops to the midpoint of the graph and then rises. Moreover, at the present time, $z =0$, we obtain $\left\{ q,r,s \right\} = \left \{-0.52,0.94,0.019 \right \}$. The value of $q=-0.52$ is close to the value for $\Lambda$CDM, $q_0\simeq-0.55$, and the values of $r$ and $s$ are also close to $1$ and $0$, respectively. After, in the limit $z \rightarrow -1$, the model gives $\left\{q,r,s \right\}  \rightarrow \left \{-1,1,0\right \}$, and the curve converges to the dS fixed point in the future.\\
In the $(s, r)$ plane, the plot starts in the region bounded by approximately $0.05< s < 0.07$ and $0.9< r \leq 1$. The trajectory is very similar to those of the $(q, r)$ plane, with a non-linear performance. We check the information given by the $(q, r)$ plane, and it is observed that the model curve converges to $\Lambda$CDM in the limit $z \rightarrow -1$. \\
We conclude that this model is compatible with $\Lambda$CDM, since in all three epochs the model is very similar to $\Lambda$CDM. It is possible that with more updated data the values for the Statefinder parameters $\left\{ q,r,s \right\}$ at present may be closer to $\left \{-0.55,1,0 \right \}$.

\subsubsection{MODEL 5. $Q_5= 3H \gamma \frac{\rho_{c}^2}{\rho}$}

 \begin{figure}[htp]
  \centering
  \label{figur}
  \subfloat[]{\label{figur322:2}\includegraphics[width=80mm]{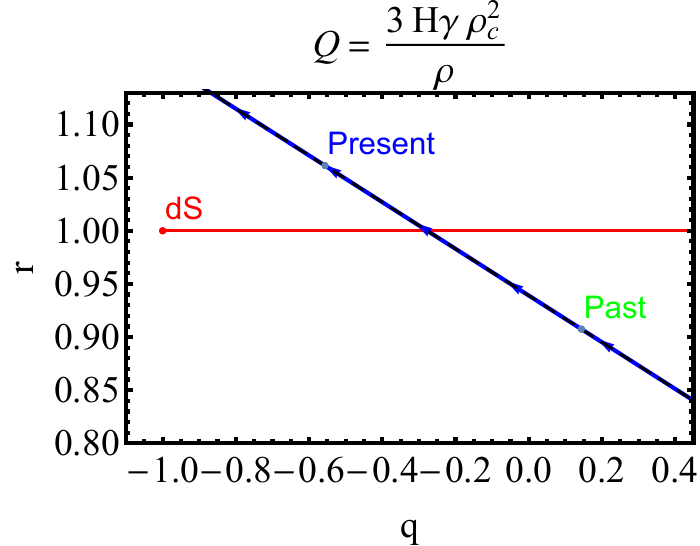}}
  \subfloat[]{\label{figur322:3}\includegraphics[width=80mm]{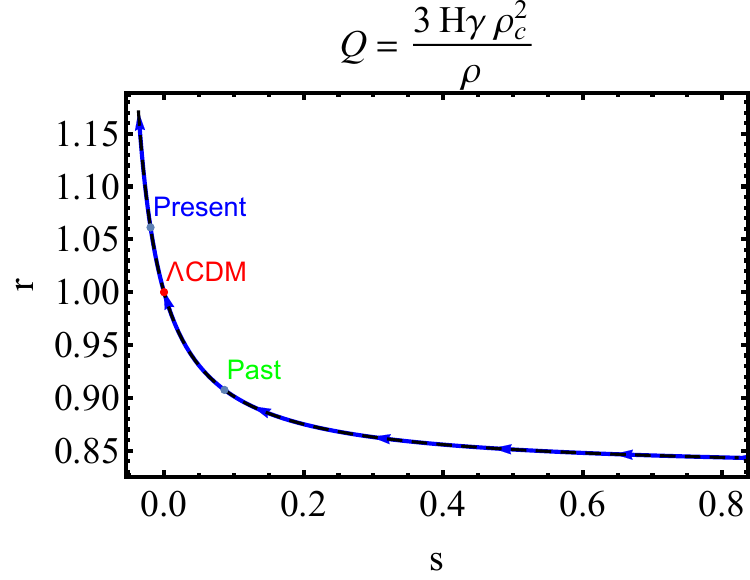}}
\caption{{\small  \textbf{Left panel}: In this figure, we have plotted the parametric curve in the $q - r$ plane. It shows the evolution for the model with the coupling function $Q_5= 3H \gamma \frac{\rho_{c}^2}{\rho}$. \textbf{Right panel}: The curve in the $s -r$ plane it is shown.}}
\end{figure}

In order to evaluate this model numerically, we will use the following values: $\Omega_{m0}= 0.320$, $\gamma=-0.038$ (recall that, we will use this value with positive sign), $w_{de}=-1$, using SNe Ia + $H_0$ + CC + BAO, from Table 3, Ref. \cite{cosmoconstrain}.
\\
In Fig.(\ref{figur322:2}), we show the curve for MODEL 5 in the $(q, r)$  plane. For early epochs, the curve starts in the lower right corner of the figure, in the region bounded by $0< q < 1$ and $r > 0$. For $z \gg 1$, the parameters $q, r$ and $s$ yield $\left\{ q,r,s \right\}  \rightarrow \left \{0.44,0.84,0.96 \right \}$, and all three parameters values deviate significantly from the values expected for the standard matter-dominated era. Only the value  for $s$ is close to that of SCDM ($s=1$).\\
We observe that the trajectory is linear and evolves in the negative direction of $q$ and positive of $r$. Before the present time, it crosses the red straight line $r=1$ corresponding to $\Lambda$CDM. At $z =0$, we obtain $\left\{ q,r,s \right\} = \left \{-0.56,1,-0.019 \right \}$. The parameter $q$ is very close to the expected value for the current time, and $r$ is exactly the value predicted by $\Lambda$CDM, while $s$ is very close to zero.
In the future, when we take the limit $z \rightarrow -1$, the model gives $\left\{ q,r,s \right\}  \rightarrow \left \{-1,1.2,-0.037 \right \}$. Then, the parameter $r$ deviates significantly from $\Lambda$CDM. The parameter $q$ is exactly the value predicted by $\Lambda$CDM and $s$ is very close to zero.
Fig. (\ref{figur322:3}) shows the $(s, r)$ plane, and the curve starts in the region bounded by $0< s < 1$ and $r > 0$. The trajectory is non-linear and evolves in the negative direction of $s$ and positive of $r$. We observe that before the current time, it crosses the fixed point $(0,1)$ corresponding to $\Lambda$CDM. The curve passes through $\Lambda$CDM but then evolves and away from it. We see that this model performs better near the present time since the values at that time are closer to $\Lambda$CDM than for other times. Then, we conclude this model is partially compatible with $\Lambda$CDM.

\subsubsection{MODEL 6. $Q_6= 3H \gamma \frac{\rho_{de}^2}{\rho}$}

\begin{figure}[htp]
  \centering
  \label{figur}
  \subfloat[]{\label{figur323:2}\includegraphics[width=80mm]{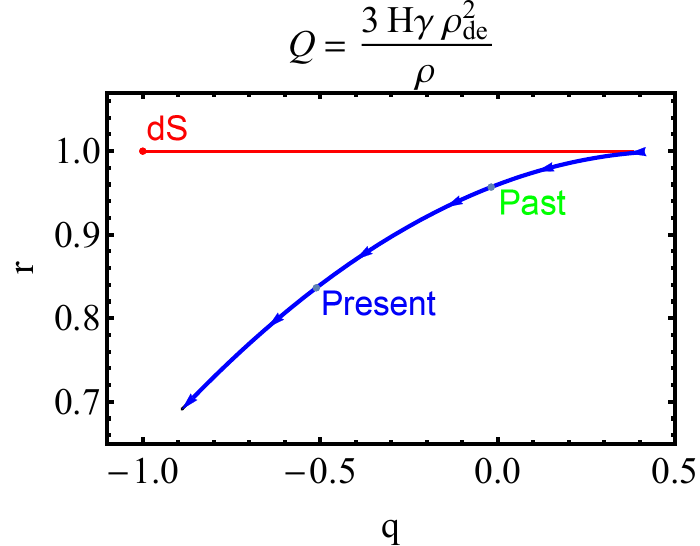}}
  \subfloat[]{\label{figur323:3}\includegraphics[width=80mm]{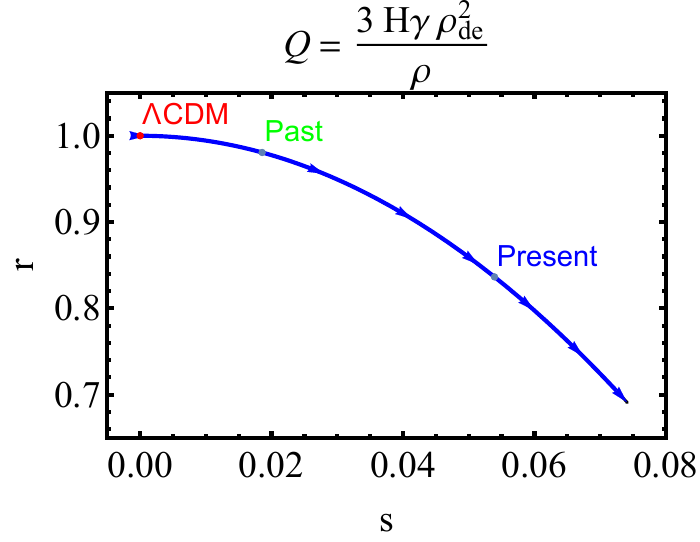}}
 \caption{{\small \textbf{Left panel}: This figure shows the evolution in the $q -r$ plane for the model with the coupling function $Q_6= 3H \gamma \frac{\rho_{de}^2}{\rho}$. \textbf{Right panel}: The graph shows the curve for this model in the $s -r$ plane.}}
\end{figure}

In order to evaluate this model numerically, we will use the following values $\Omega_{m0}= 0.326$, $\gamma=-0.08$, $w_{de}=-1$, using SNe Ia + $H_0$ + CC + BAO, from Table 3, Ref. \cite{cosmoconstrain}.
\\
Fig.(\ref{figur323:2}) shows the evolution for MODEL 6 in the $(q, r)$ plane. For early epochs, the trajectory  starts in the upper right corner of the figure, in the region bounded by $0< q < 1$ and $0<r\leq 1$. In this epoch, when $z \gg 1$ the parameters $q, r$ and $s$ yield $\left\{ q,r,s \right\}  \rightarrow \left \{0.5,1,0 \right \}$, and the values of $q$, $r$ and $s$ agree with the matter-dominated standard era.
The trajectory behaves much like MODEL 2, where the interaction also depends on the DE energy density. We see that in this case, the trajectory is curved and evolves in the negative $q$ and negative $r$ direction.\\
At the past the curve moves away from the red straight line $r=1$ corresponding to  $\Lambda$CDM, and then moves forward. At $z =0$, we obtain $\left\{ q,r,s \right\} = \left \{-0.51,0.84,0.05 \right \}$.  
The parameter $q$ is very close to the expected value for the present time, while $r$ is far away to the value predicted by $\Lambda$CDM, and $s$ is very close to zero. After, the trajectory moves away significantly from the dS fixed point, because in the limit $z \rightarrow -1$, the model gives $\left\{ q,r,s \right\} \rightarrow \left \{-0.89,0.69,0.074 \right \}$.
Fig. (\ref{figur323:3}) shows the $(s, r)$ plane, where the trajectory starts in the region bounded by $0< s < 1$ and $0<r\leq 1$ and has  non-linear behaviour, evolves in the positive direction of $s$ and negative of $r$. We observe that it starts at the fixed point $(0,1)$ corresponding to $\Lambda$CDM, and then evolves to current time and diverges away. It is concluded that this model is partially compatible with $\Lambda$CDM.

\subsection{Models with a change of direction of energy transfer}

\subsubsection{MODEL 7. $Q_7= q(\alpha \dot{\rho}_{c} + 3 \beta H\rho_{c})$}

For $\alpha=0$, it yields $Q_7= 3 \beta q H\rho_{c}$.

\begin{figure}[htp]
  \centering
  \label{figur}
  \subfloat[]{\label{figur331:1}\includegraphics[width=80mm]{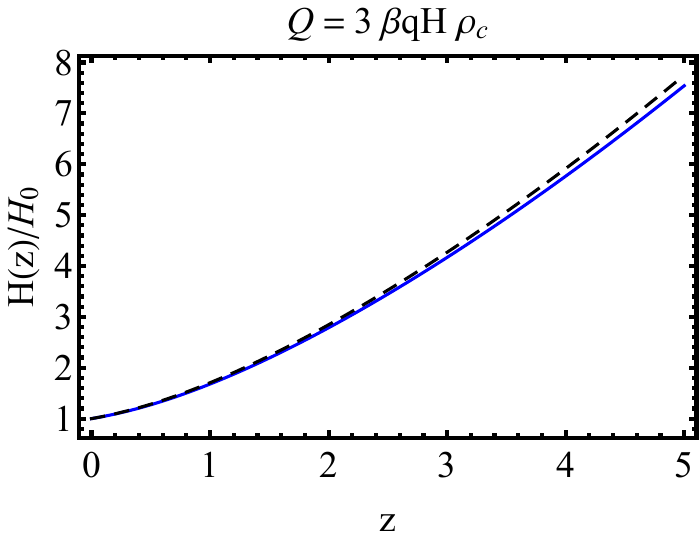}}
  \subfloat[]{\label{figur331:2}\includegraphics[width=80mm]{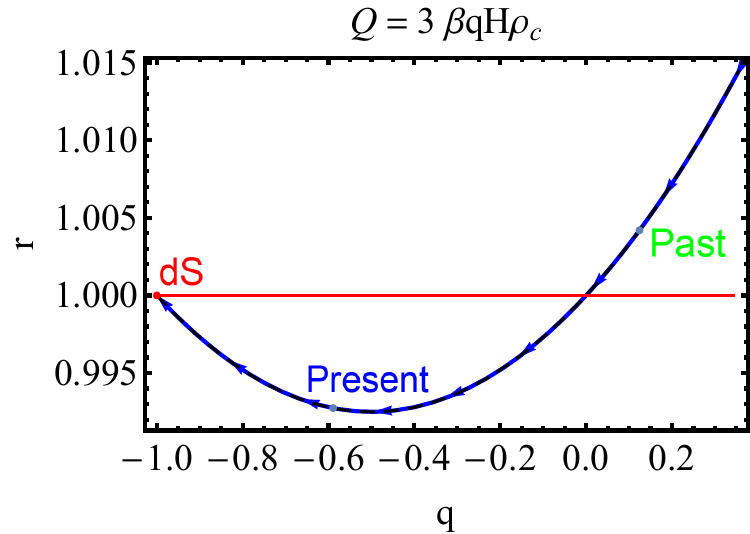}}
   \\
  \subfloat[]{\label{figur331:3}\includegraphics[width=80mm]{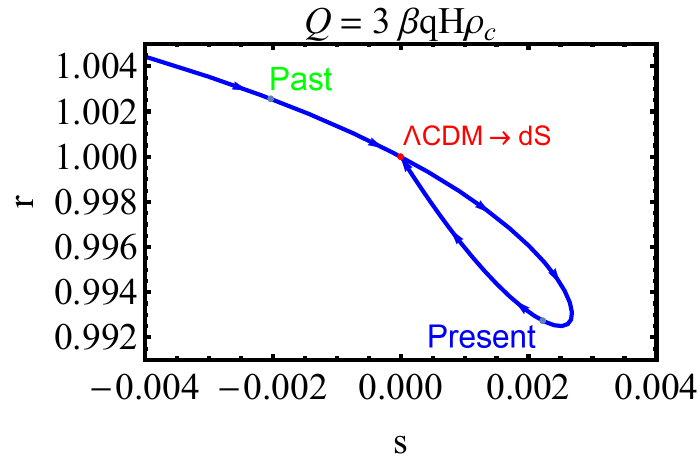}}
  \caption{{\small Figure (a) shows the dimensionless Hubble rate $H(z)/H_0$ as a function of redshift $z$ for the model with the coupling function $Q_7= 3 \beta q H\rho_{c}$. We represent the interaction with a dashed line, and also the Hubble rate of the $\Lambda$CDM model with the solid line. For this model, Figures (b) and (c) show the curves in the $q - r$ and $s - r$ planes, respectively.}}
\end{figure}

We will evaluate this model numerically using the following values $\Omega_{c0}= 0.2738$, $\Omega_{b0} = 0$, $\beta =-0.01$, $w_{de}=-1$, of TABLE I from Ref. \cite{hao2011cosmological}.
\\
For MODEL 7 in the $(q, r)$ plane, we observe from Fig.(\ref{figur331:2}) that the trajectory is non-linear and starts in the upper right corner, in the region bounded by $0< q < 1$ and $0< r<1$. The curve evolves to the negative direction of $q$ and negative of $r$.
For $z \gg 1$, the parameters $q, r$ and $s$ behaves as $\left\{ q,r,s \right\}  \rightarrow \left \{0.51,1,1 \right \}$, and $q$ and $r$ correspond to the standard matter-dominated era, and $s$ has the value corresponding to SCDM. 
Before the present time, the trajectory crosses the red straight line $r=1$ corresponding to $\Lambda$CDM and evolves.
At $z =0$, we obtain $\left\{ q,r,s \right\} = \left \{-0.59,0.99,0.002 \right \}$, which are close to the values expected for $\Lambda$CDM. In the limit $z \rightarrow -1$, the model gives $\left\{ q,r,s \right\}  \rightarrow \left \{-1,1,0 \right \}$, which correspond exactly with to the dS fixed point in the future. Thus, the model converges to $\Lambda$CDM.
\\
In the $(s, r)$ plane, Fig.(\ref{figur331:3}) shows that the curve starts in the upper left corner, in the region bounded by $0 < s$ and $r > 1$. 
The trajectory is non-linear, passing through the point $(0,1)$ corresponding to $\Lambda$CDM before the present time. After the present time, the curve generates a twist and closes itself, converging to the point $(0,1)$. 
We corroborate the information provided by the $(q, r)$ plane; the trajectory of the model converges to $\Lambda$CDM. As a consequence, we conclude that this model is compatible with $\Lambda$CDM, because the values of Statefinder parameters are very close to the values predicts for $\Lambda$CDM, in general, and in the future, the model evolves towards $\Lambda$CDM.

\subsubsection{MODEL 8. $Q_8= q(\alpha \dot{\rho}_{tot} + 3 \beta H\rho_{tot})$}

%In the case for $\alpha=0$, it yields $Q_8= 3 \beta q H\rho_{tot}$.

\begin{figure}[htp]
  \centering
  \label{figur}
  \subfloat[]{\label{figur332:2}\includegraphics[width=80mm]{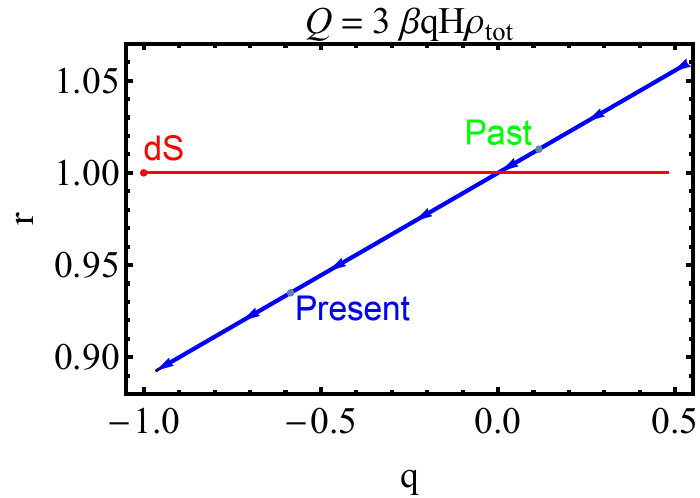}}
  \subfloat[]{\label{figur332:3}\includegraphics[width=80mm]{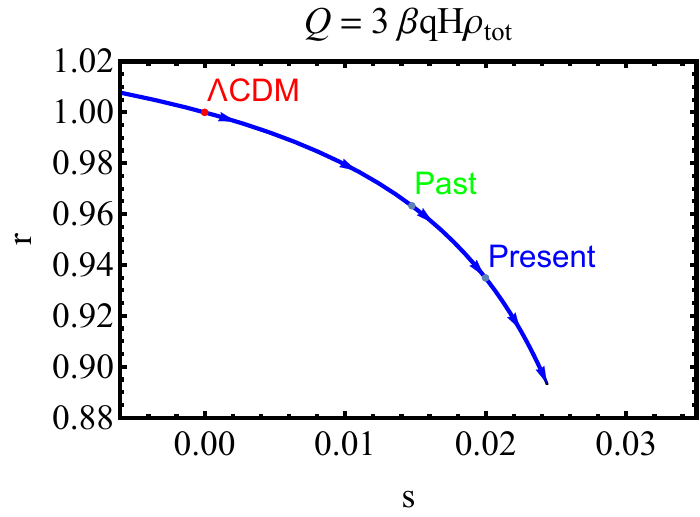}}
  \caption{{\small Figures (a) and (b) show the evolution for the model with the coupling function $Q_8= 3 \beta q H\rho_{tot}$. \textbf{Left panel}: Figure shows the $q - r$ parametric curve.
\textbf{Right panel}: Figure shows the $s - r$ parametric curve.}}
 \end{figure}

In order to evaluate this model numerically, we use the following values: $\Omega_{c0}= 0.2764$, $\Omega_{b0} = 0$, $\beta =-0.0247$, $w_{de}=-1$, Table I from Ref. \cite{hao2011cosmological}.
\\
Fig. (\ref{figur332:2}) shows the evolution for MODEL 8 in the $(q, r)$ plane. For early epochs,  the trajectory  starts in the upper right corner of the figure, in the region bounded by $0< q < 1$ and $0<r \lesssim1.1$. In this epoch, when $z \gg 1$ the Statefinder parameters $q, r$ and $s$ behave as $\left\{ q,r,s \right\}  \rightarrow \left \{0.52,1.1,1 \right \}$, and the values of $q$ and $r$ agree with the matter-dominated standard era, while $s$ deviates significantly from the value $s=0$ for $\Lambda$CDM. The value $s=1$ corresponds to SCDM, being very similar to MODEL 7 in this epoch.
We see that in this case, the trajectory is linear and evolves in the negative $q$ and negative $r$ direction. 
At the past, the curve moves away from the red straight line $r=1$ corresponding to  $\Lambda$CDM, and evolves. At $z =0$, we obtain $\left\{ q,r,s \right\}  = \left \{-0.59,0.93,0.02 \right \}$, which are close to the values expected for $\Lambda$CDM. After, the trajectory moves away significantly from the dS fixed point, because in the limit $z \rightarrow -1$, we obtain $\left\{ q,r,s \right\}  \rightarrow \left \{-0.96,0.89,0.024 \right \}$.
\\
Fig. (\ref{figur332:3}) shows the $(s, r)$ plane, where the trajectory starts in the region bounded by $0< s < 1$ and $0<r \lesssim1.1$ and has  non-linear behaviour, evolves in the positive direction of $s$ and negative of $r$. We observe that it starts at the fixed point $(0,1)$ corresponding to $\Lambda$CDM, and then evolves to the current time and diverges away. It is concluded that this model is partially compatible with $\Lambda$CDM.

\subsubsection{MODEL 9. $Q_9= q(\alpha \dot{\rho}_{de} + 3 \beta H\rho_{de})$}

%When $\alpha=0$ the interaction reduces to $Q_9= 3 \beta q H \rho_{de}$.

\begin{figure}[htp]
  \centering
  \label{figur}
  \subfloat[]{\label{figur333:2}\includegraphics[width=80mm]{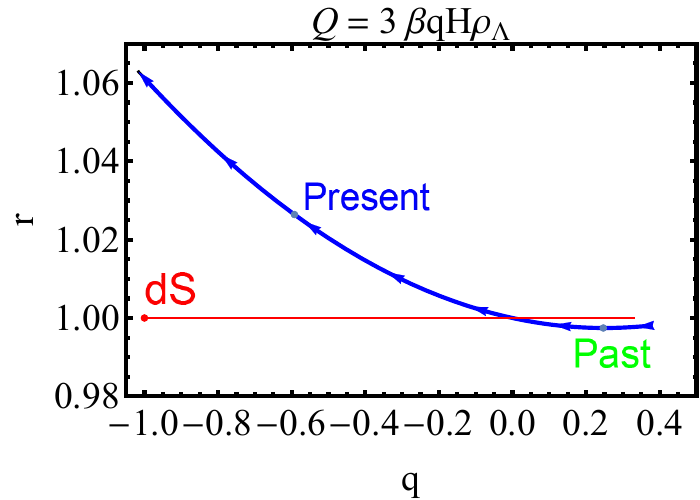}}
  \subfloat[]{\label{figur333:3}\includegraphics[width=80mm]{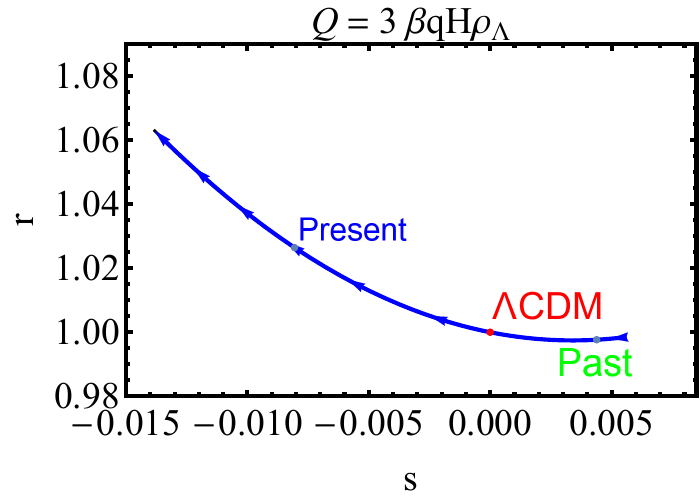}}
    \caption{{\small Figures (a) and (b) show the evolution for the model with the coupling function $Q_9= 3 \beta q H \rho_{de}$. \textbf{Left panel}: Figure shows the $q - r$ parametric curve.
\textbf{Right panel}: Figure shows the $s - r$ parametric curve.}}
\end{figure}

The values used for to evaluate this model numerically are: $\Omega_{c0}=0.2717$, $\Omega_{b0} = 0$, $\beta = 0.0136$, $w=-1$, from TABLE I from Ref.\cite{hao2011cosmological}. As in this Ref., the author used $w_{de}=-1$, the "$\Lambda$" subscript that we use in the graphs is replacing in the equations by the subscript "de".
\\
We see in Fig.(\ref{figur333:2}) the curve for MODEL 9 in the $(q, r)$ plane. The trajectory starts in the lower right corner of the figure, in the region bounded by $0< q < 1$ and $r > 0$. For $z \gg 1$, the parameters $q, r$ and $s$ behaves as $\left\{ q,r,s \right\}  \rightarrow \left \{0.5,1.03,0.007 \right \}$, and all three parameter values are very close from the values expected for the standard matter-dominated era. We observe that the trajectory is non-linear and evolves in the negative direction of $q$ and positive of $r$. Before the present time ($z=0$), it crosses the red straight line $r=1$ corresponding to $\Lambda$CDM. At the present, we obtain $\left\{ q,r,s \right\} = \left \{-0.59,1.02,-0.008 \right \}$. The parameters are close to the expected value predicted by $\Lambda$CDM for the present time.
In the future, $z \rightarrow -1$, the model gives $\left\{ q,r,s \right\} \rightarrow \left \{-1.02,1.06,-0.014 \right \}$. Then, the trajectory moves away significantly from the dS fixed point. \\
Fig. (\ref{figur333:3}) shows the $(s, r)$ plane, and the curve starts in the region bounded by $0< s < 1$ and $r > 0$. The trajectory is non-linear and evolves in the negative direction of $s$ and positive of $r$. We observe that before present time, it crosses the fixed point $(0,1)$. The curve passes through $\Lambda$CDM, but evolves and away from it. We say that this model is partially compatible with $\Lambda$CDM.

\subsection{Interaction involving derivatives}
%\\
\subsubsection{MODEL 10. $Q_{10}=3\alpha H (\rho_{de}'+\rho_{c}')$}

\begin{figure}[htp]
  \centering
  \label{figur}

  \subfloat[]{\label{figur10:1}\includegraphics[width=80mm]{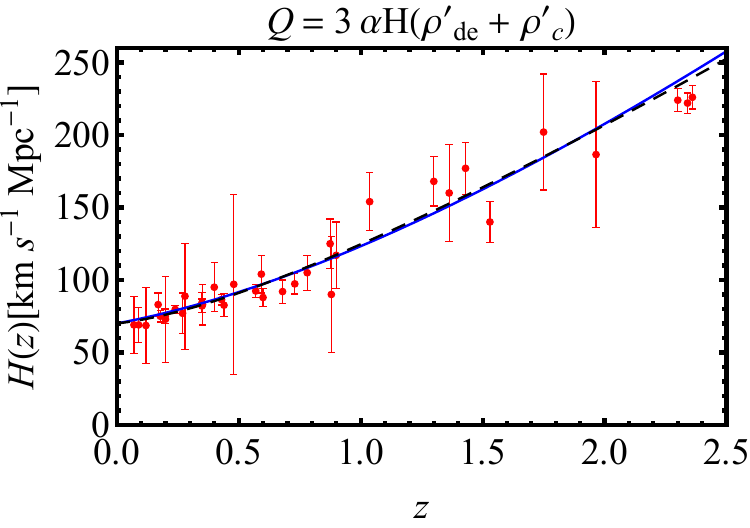}}
   \\
  \subfloat[]{\label{figur10:2}\includegraphics[width=80mm]{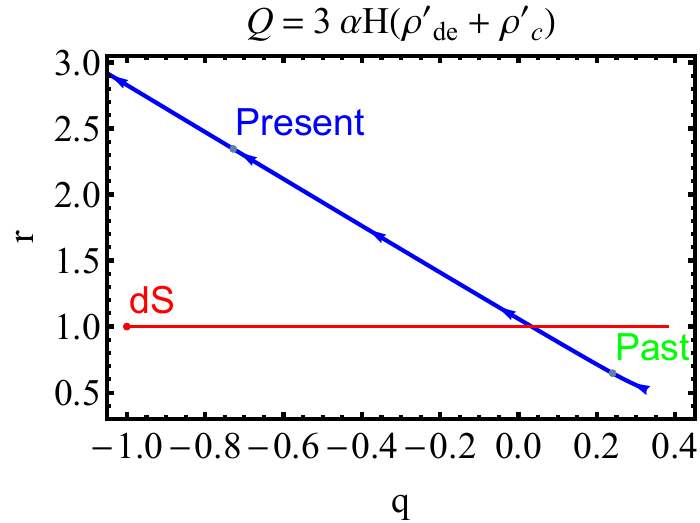}}
  \subfloat[]{\label{figur10:3}\includegraphics[width=80mm]{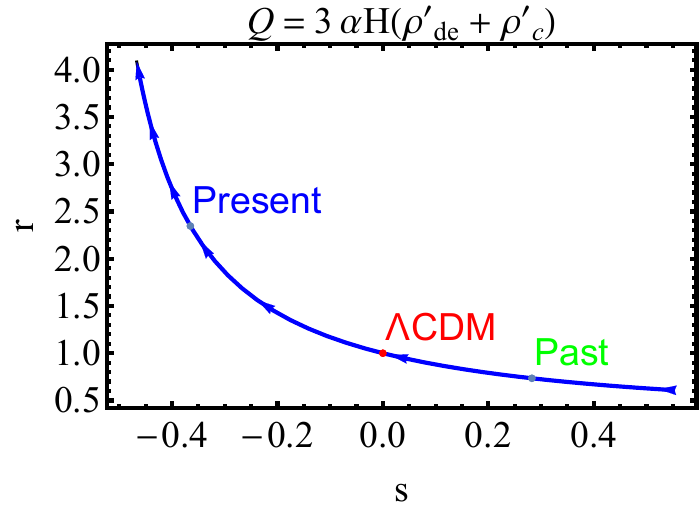}}
 \caption{{\small In Figure (a) it is shown of the Hubble rate $H(z)$ for the model with the coupling function $Q_{10}=3\alpha H (\rho_{de}'+\rho_{c}')$. We have used  $H_0=70 {~}\mathrm{km~s^{-1}~Mpc^{-1}}$, from Ref. \cite{Grandon:2018hom}. Figures (b) and (c) show the parametric curves in the $q - r$ and $s - r$ planes, respectively.}} 
\end{figure}
In order to evaluate this model numerically, we will use the values: $h=0.70$, $\Omega_{c}= 0.37$, $\Omega_{b}=0.045$, $\gamma=0.071$, $w_{de}= -1.4$, $\alpha = -0.15$, extracted from Ref. \cite{Grandon:2018hom}, SNIa+$H(z)$+BAO+$f_{gas}$, Table IV, Column B. We have preferred to use the values in column B, because in column C, although the CMB measurements are included, we have $\alpha=0.0\pm0.1$, so there is almost no interaction.
\\
In Fig.(\ref{figur10:2}) we see the curve for MODEL 10 in the $(q, r)$ plane. For early times, the trajectory starts in the lower right corner of the figure, in the region bounded by $0< q < 1$ and $r > 0$. For $z \gg 1$, the parameters $q, r$ and $s$ behave as $\left\{ q,r,s \right\} \rightarrow \left \{0.38, 0.68,0.86 \right \}$. All the three parameter values deviate significantly from the values expected for the standard matter-dominated era. The value of $s=0.86$ is close to SCDM ($r=1, s=1$), but lower than the other models that evolve in this way. We observe that the trajectory is linear and evolves in the negative direction of $q$ and positive of $r$. Before the present time ($z=0$) it crosses the red straight line $r=1$ which  corresponds to $\Lambda$CDM. At $z =0$, we obtain $\left\{ q,r,s \right\} = \left \{-0.73,2.35,-0.37 \right \}$, and then all the three parameters values deviate significantly from the values expected for $\Lambda$CDM. In the future, for $z \rightarrow -1$, it yields $\left\{ q,r,s \right\} \rightarrow \left \{-1.7,4.1,-0.47 \right \}$. Then, all the parameters deviate significantly from $\Lambda$CDM.
\\
Fig. (\ref{figur10:3}) shows the $s - r$ plane, and the curve starts in the region bounded by $0< s < 1$ and $r > 0$. The trajectory is non-linear and evolves in the negative direction of $s$ and positive of $r$. The plot in the $s - r$ plane shows that in the first place, the curve passes through $\Lambda$CDM. After, it passes through the point associated with the present time, and then it deviates significantly in the future. We say that this model is incompatible with $\Lambda$CDM, since the Statefinder values deviate significantly from the expected values in all three epochs.

\subsection{Parameterized interactions}

\subsubsection{MODEL 11. $f(\tilde{r})=1$}

\begin{figure}[htp]
  \centering
  \label{figur}

  \subfloat[]{\label{figur11:1}\includegraphics[width=80mm]{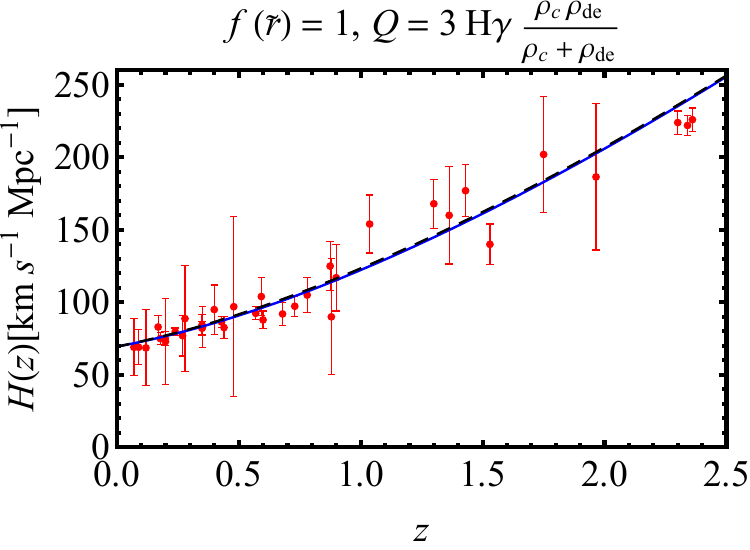}}
  \subfloat[]{\label{figur11:2}\includegraphics[width=80mm]{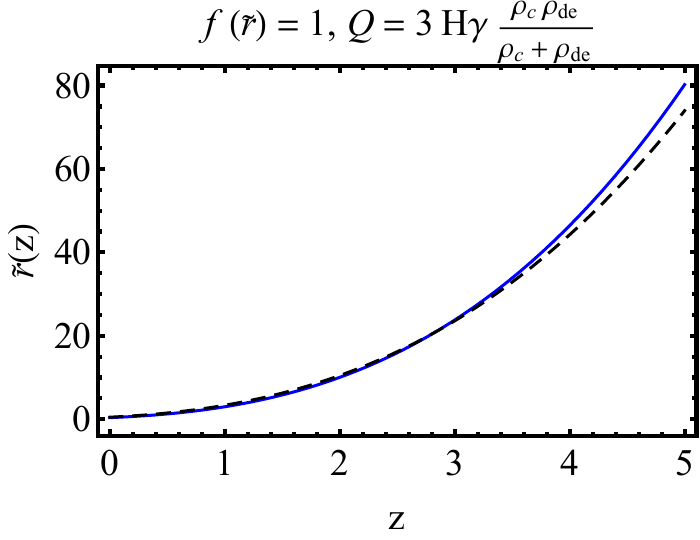}}
  \\
  \subfloat[]{\label{figur11:3}\includegraphics[width=80mm]{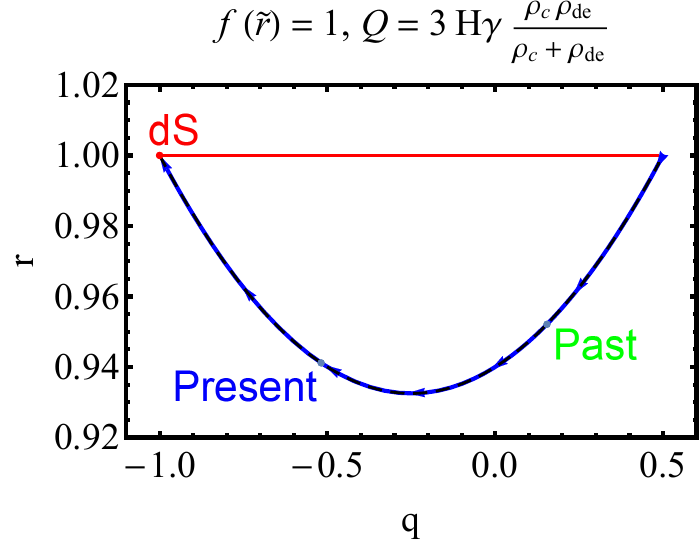}}
  \subfloat[]{\label{figur11:4}\includegraphics[width=80mm]{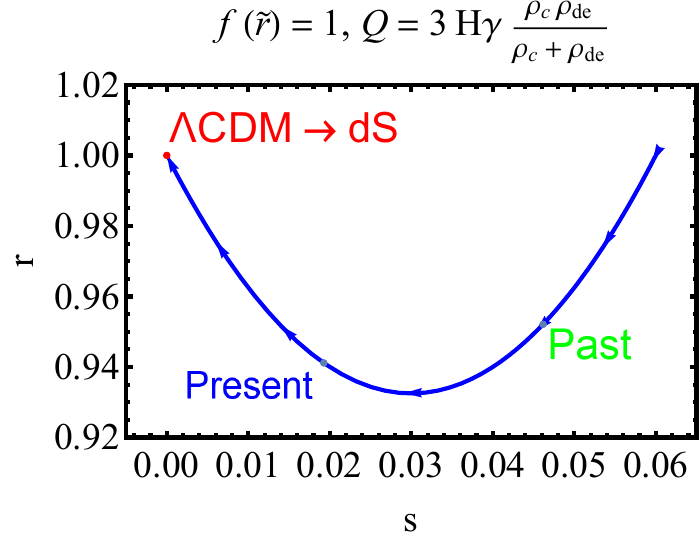}}
  
\caption{{\small In Figure (a) it is shown the Hubble rate $H(z)$ as a function of redshift $z$ for the model with function  $f(\tilde{r})=1$. We have used  $H_0=69.44 {~}\mathrm{km~s^{-1}~Mpc^{-1}}$, from \cite{cosmoconstrain}. For the same model, Figure (b) shows the evolution of the ratio $\tilde{r}$, where we have represented the interacting model (black dashed line), which overlaps with $\Lambda$CDM (blue solid line).  Figures (c) and (d) show the parametric curves in the $q - r$ and $s - r$ planes, respectively.}}  
 \end{figure}

We have used the values $\Omega_{m0}= 0.321$, $\gamma=-0.06$, $w_{de}=-1$, using SNe Ia + $H_0$ + CC + BAO, from Table 3 \cite{cosmoconstrain}. This values are the same that for MODEL 4 to which this parameterization corresponds. Figs. (\ref{figur11:3}) and (\ref{figur11:4}) show that we have obtained exactly the same parametric plots that in the MODEL 4, given by Figs. (\ref{figur321:2}) and (\ref{figur321:3}). Furthermore, present and asymptotic values of Statefinder parameters become the same, so the Statefinder analysis is the same, see Table (\ref{statevalues}). On the other hand, we have included the plot of the ratio between the DM and DE densities against $z$. We have corroborated what the authors of the reference \cite{cosmoconstrain} claim, that for $\gamma<0$, the cosmic coincidence problem may be considerably alleviated. At early times, DM dominates over DE, if $z \gg 1$ then $\tilde{r}(z) \rightarrow \infty$. If $z \rightarrow -1$ then $\tilde{r}(z) \rightarrow 0$, thus, DE dominates over DM in the far future. For negative value of the coupling, the ratio $\tilde{r}(z)$ yields the order of 1 earlier than in the $\Lambda$CDM model ($\gamma= 0$).

\subsubsection{MODEL 12. $f(\tilde{r})=\frac{1}{\tilde{r}}$}

\begin{figure}[htp]
  \centering
  \label{figur}
   \subfloat[]{\label{figur12:3}\includegraphics[width=80mm]{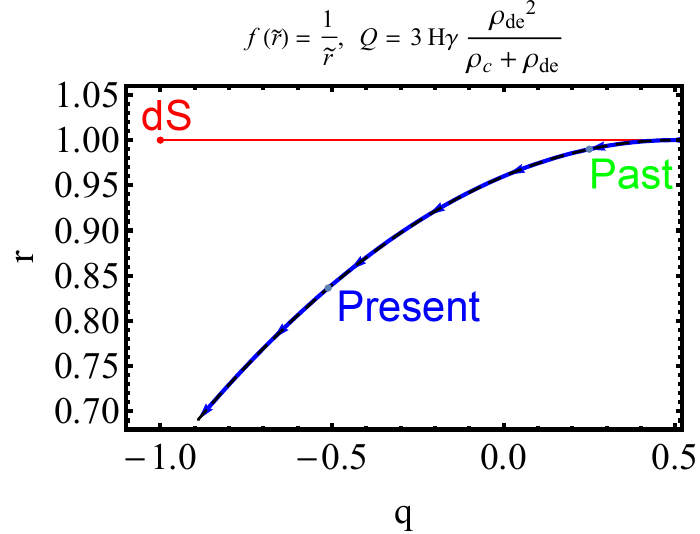}}
  \subfloat[]{\label{figur12:4}\includegraphics[width=80mm]{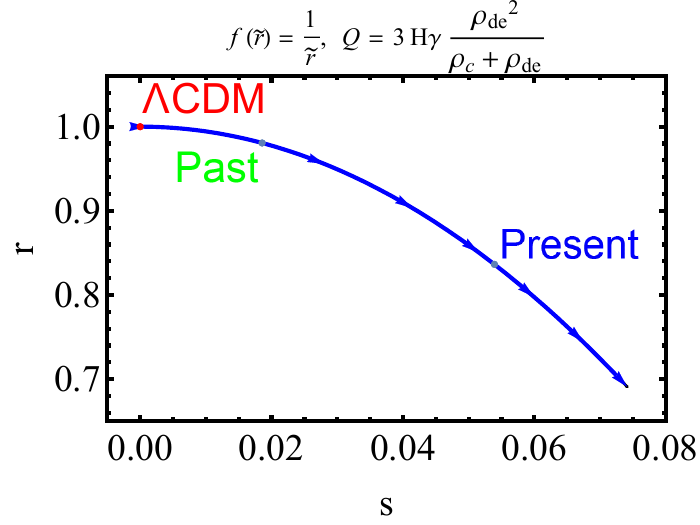}}
  \caption{{\small In the figures we show the evolution for the model with function $f(\tilde{r})=\frac{1}{\tilde{r}}$. \textbf{Left panel}: Figure shows the $q - r$ parametric curve. \textbf{Right panel}: Figure shows the $s - r$ parametric curve.}} 
  \end{figure}
Here we use the values $\Omega_{m0}= 0.326$, $\gamma=-0.08$, $w_{de}=-1$, considering SNe Ia + $H_0$ + CC + BAO, from Table 3, Ref. \cite{cosmoconstrain}. As in the previous case, this model is equivalent to MODEL 6, thus, we have obtained exactly the same parametric plots and present and asymptotic values for the Statefinder parameters. See Figs.(\ref{figur12:3}) and (\ref{figur12:4}) compared with Figs.(\ref{figur323:2}) and (\ref{figur323:3}). From Table (\ref{statevalues}), we conclude that the Statefinder analysis becomes the same.

\subsubsection{MODEL 13. $f(\tilde{r})=\tilde{r}$}

\begin{figure}[htp]
  \centering
  \label{figur}
  \subfloat[]{\label{figur13:3}\includegraphics[width=80mm]{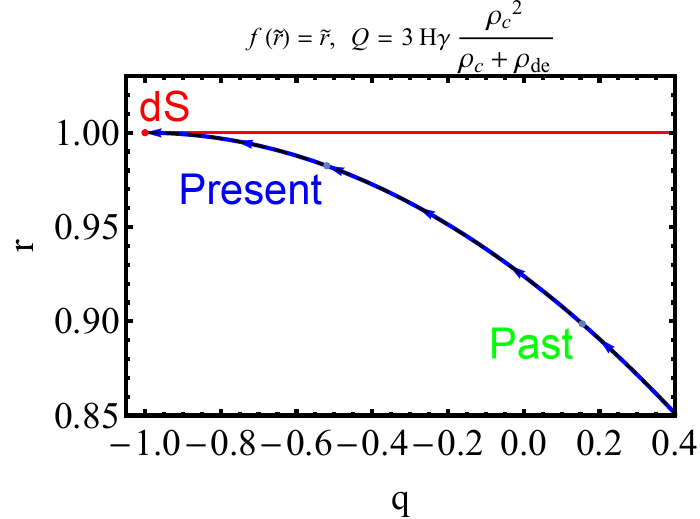}}
  \subfloat[]{\label{figur13:4}\includegraphics[width=80mm]{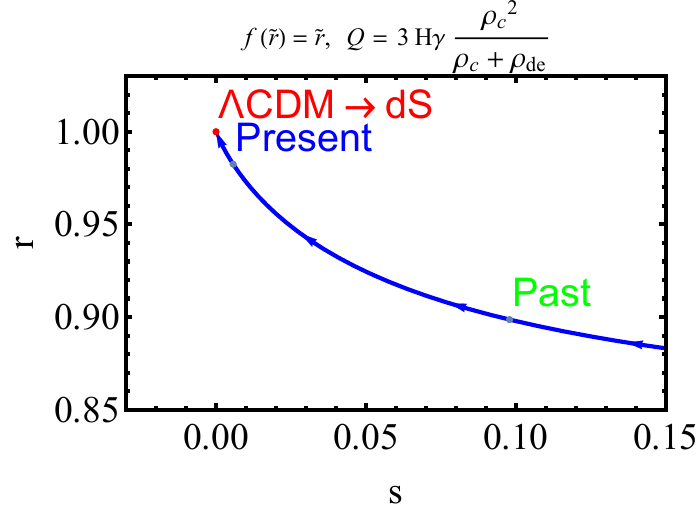}}
\caption{{\small For the model with function $f(\tilde{r})=\tilde{r}$, the parametric curves in the $q - r$ and $s - r$ planes are shown.}} 
\end{figure}

This model is equivalent to MODEL 5, so we will use the following values: $\Omega_{m0}=0.320$, $\gamma=-0.038$, $w_{de}=-1$, using SNe Ia + $H_0$ + CC + BAO, from Table 3, Ref. \cite{cosmoconstrain}.
\\
In Fig.(\ref{figur13:3}), we see the curve for MODEL 13 in the $(q, r)$ plane. The trajectory starts in the lower right corner of the figure, in the region bounded by $0< q < 1$ and $0<r <1$. We calculate the parameters $q, r$ and $s$ at $z \gg 1$, then  $\left\{ q,r,s \right\} \rightarrow \left \{0.44,0.84,0.96 \right \}$, and then all three parameter values deviate significantly  from the values expected for the standard matter-dominated era. The value of $s=0.96$ is very close to SCDM.\\
We observe that the trajectory is non-linear and evolves in the negative direction of $q$ and positive of $r$. At the present $z =0$, we obtain  $\left\{ q,r,s \right\}  = \left \{-0.52,0.98,0.006 \right \}$. The parameters are close to the values predicted by $\Lambda$CDM for the present time. After the present time it approaches to the red straight line $r=1$ corresponding to $\Lambda$CDM. 
We verify this, since in the future, in the limit $z \rightarrow -1$, the model gives $\left\{ q,r,s \right\}  \rightarrow \left \{-1,1,0 \right \}$. Then, the trajectory converges to the fixed dS point. \\
Fig. (\ref{figur13:4}) shows the $(s, r)$ plane, the curve starts in the region bounded by $0< s < 1$ and $r > 0$. The trajectory is non-linear and evolves in the negative direction of $s$ and positive of $r$. We notice that at late times, it converges to the fixed point $(0,1)$ associated to $\Lambda$CDM. Hence, this model becomes partially compatible with $\Lambda$CDM.

\subsubsection{MODEL 14. $f(\tilde{r})=1+\frac{1}{\tilde{r}}$}

\begin{figure}[htp]
  \centering
  \label{figur}
   \subfloat[]{\label{figur14:3}\includegraphics[width=80mm]{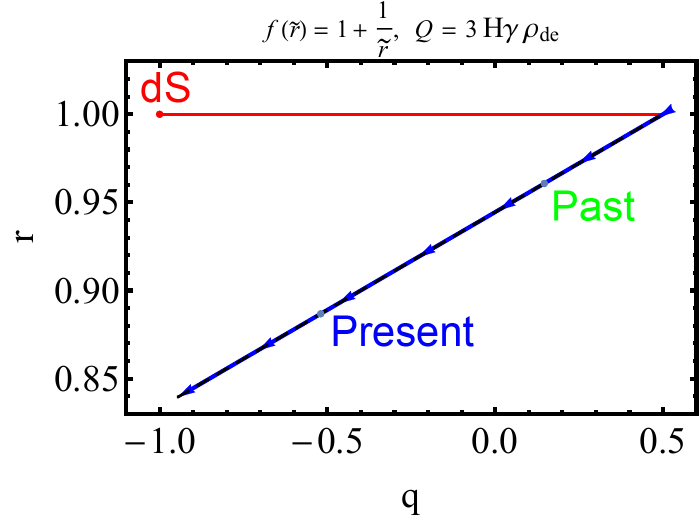}}
  \subfloat[]{\label{figur14:4}\includegraphics[width=80mm]{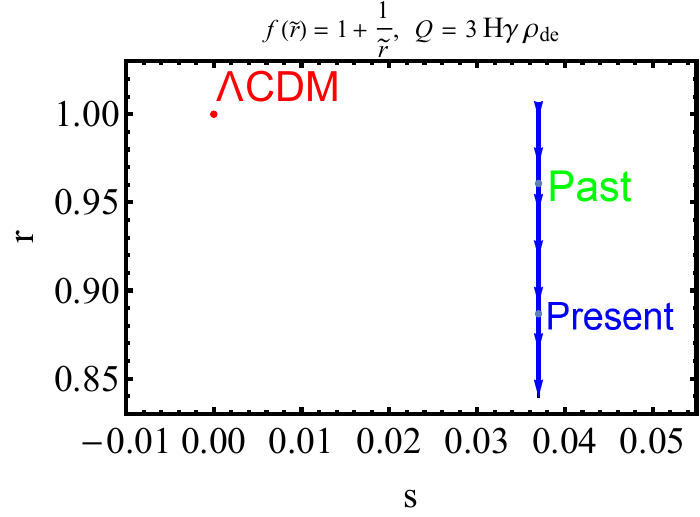}}
  
\caption{{\small For the model with function $f(\tilde{r})=1+\frac{1}{\tilde{r}}$, the parametric curves in the $q - r$ and $s - r$ planes are shown.}}   
\end{figure}

In this case, we have used the values: $\Omega_{m0}=0.320$, $\gamma=-0.037$, $w_{de}=-1$, considering SNe Ia + $H_0$ + CC + BAO, from Table 3, Ref. \cite{cosmoconstrain}.\\
In the high redshift limit, $z \gg 1$, the asymptotic values of $q, r$ and $s$ behave as $\left\{ q,r,s \right\} \rightarrow \left \{0.5,1,0.037 \right \}$. For the present epoch $z =0$, we obtain $\left\{ q,r,s \right\} = \left \{-0.52,0.89,0.037 \right \}$. In the limit $z \rightarrow -1$, the model gives $\left\{ q,r,s \right\} \rightarrow \left \{-0.94,0.84,0.037 \right \}$.
This model is equivalent to MODEL 2, and its behaviour in the $(q, r)$ and $(s, r)$ plane becomes the same, but as for the parameterized model we have used updated data. We can compare Figs. (\ref{figur_DE:2}) and (\ref{figur_DE:3}) with Figs. (\ref{figur14:3}) and (\ref{figur14:4}), respectively. We notice that the trajectory in the $(s, r)$ plane evolves in the negative direction of $r$ while holding fixed $s=0.037$, but in MODEL 2 the trajectory gets lower by $s=0.02$. 

\subsubsection{MODEL 15. $f(\tilde{r})=1+\tilde{r}$}

\begin{figure}[htp]
  \centering
 
  \label{figur}
  \subfloat[]{\label{figur15:3}\includegraphics[width=80mm]{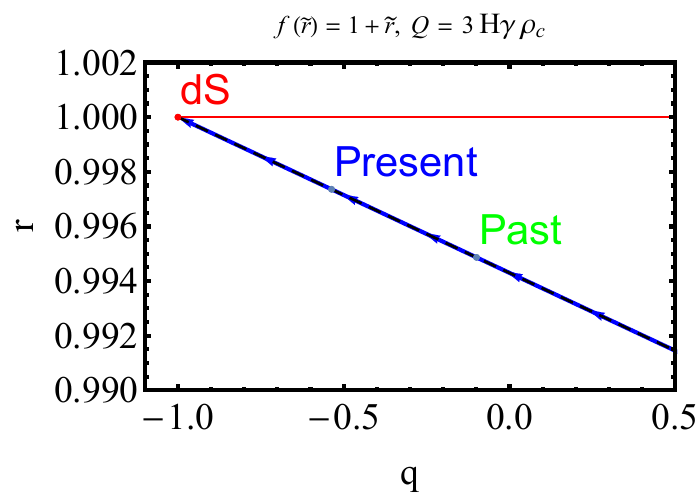}}
  \subfloat[]{\label{figur15:4}\includegraphics[width=80mm]{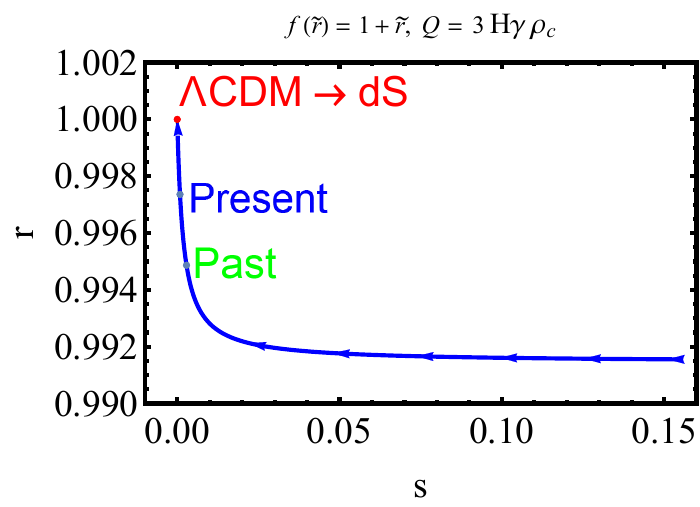}}
  
\caption{{\small We show the parametric curves in the $q - r$ and $s - r$ planes for the model with parametric function $f(\tilde{r})=1+\tilde{r}$.}}  
\end{figure}
This model is equivalent to MODEL 1, but, in this case, we have used the values $\Omega_{m0}=0.309$, $\gamma=-0.0019$, $w_{de}=-1$, considering SNe Ia + $H_0$ + CC + BAO, from Table 3, Ref. \cite{cosmoconstrain}. 
For early times, i.e. $z \gg 1$, the parameters $q, r$ and $s$ behave as$\left\{ q,r,s \right\}  \rightarrow \left \{0.5,0.99,1\right \}$. The value of $s=1$ corresponds to SCDM. We recover the standard matter-dominated era due the values obtained for $r$ and $q$.\\
At the present time, we get $\left\{ q,r,s \right\}  = \left \{-0.54,0.991,0.001 \right \}$. In the limit $z \rightarrow -1$, the model gives $\left\{ q,r,s \right\}  \rightarrow \left \{-1,1,0 \right \}$.\\
We compare Figs.(\ref{figurc:2}) and (\ref{figurc:3}) with Figs. (\ref{figur15:3}) and (\ref{figur15:4}), respectively. It is important to point out that for MODEL 1, the trajectories in the $q- r$ and $s- r$ planes do not converge to the point $(-1,1)$ and $(0,1)$, respectively. 
However, the trajectories for MODEL 15 in both planes do converge to these points.
This model has similar behavior to MODEL 1, but for the parameterized model we have used updated data and $w=-1$. This may explain the convergence in the future to $\Lambda$CDM, since in MODEL 1 we have used $w \neq -1$. In this way, we say that this model is compatible with $\Lambda$CDM.

\subsection{Self-interaction between DM}

\subsubsection{MODEL 16. {\small Symmetric model. $Q_m=-3H\alpha\rho_m, ~~~ Q_x=-3H\beta\rho_x$}}

\begin{figure}[htp]
  \centering
  \label{figur}

  \subfloat[]{\label{figur16:1}\includegraphics[width=80mm]{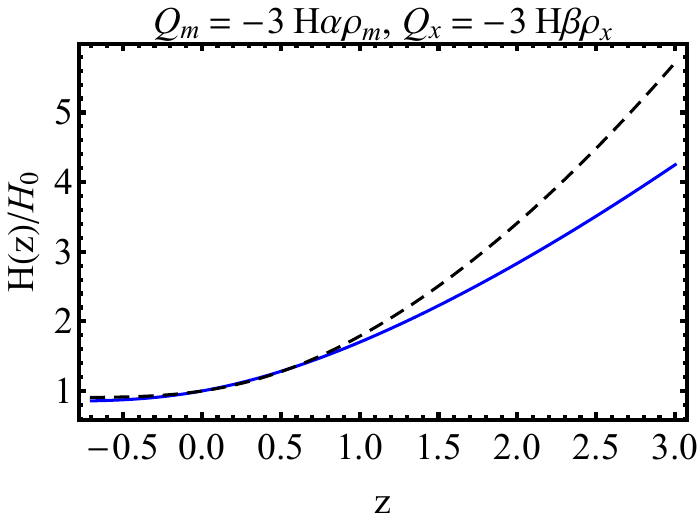}}
   \\
  \subfloat[]{\label{figur16:2}\includegraphics[width=80mm]{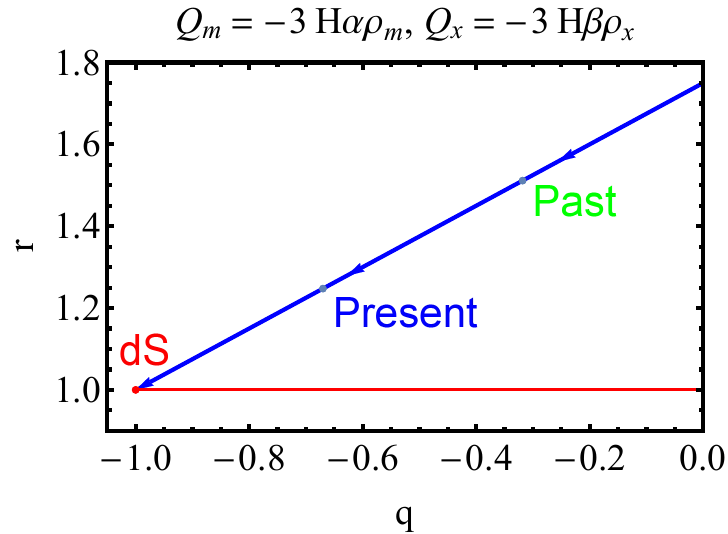}}
  \subfloat[]{\label{figur16:3}\includegraphics[width=80mm]{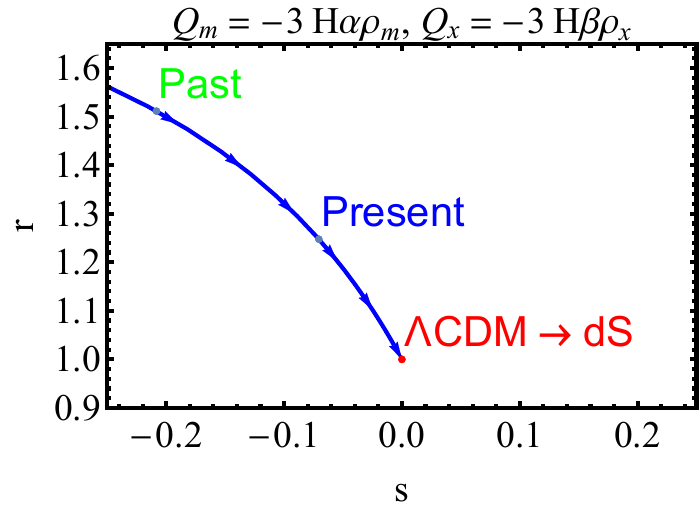}}
\caption{{\small In Figure (a), it is shown the dimensionless Hubble rate $H(z)/H_0$ as a function of redshift $z$ for the Symmetric model, $Q_m=-3H\alpha\rho_m, ~~~ Q_x=-3H\beta\rho_x$. It is represented the self-interaction (dashed line), as also, the Hubble rate of $\Lambda$CDM model (solid line). For this model, Figures (b) and (c) show the parametric curves in the $q - r$ and $s - r$ planes, respectively.}}
\end{figure}

Given the symmetry of MODEL 16, it is evaluated using the values extracted from Ref. \cite{Cardenas:2020nny}, $\Omega_m= \Omega_x=0.11 \pm 0.03$, and $\alpha=\beta=0.25 \pm0.15$. Furthermore, if we evaluate numerically the Symmetric and Asymmetric models using $\Omega_{m}=0.27$, and setting $\alpha=\beta \approx 0.001$ (we cannot use $\alpha$ and $\beta$ null), we get that the model is similar to $\Lambda$CDM.\\
In Fig. (\ref{figur16:1}), we show the dimensionless Hubble rate against $z$, and we represent the self-interaction (dashed line) overlap to the Hubble rate for the $\Lambda$CDM model (solid line). We have calculated using the above values that for $z < 0.56$ both functions are almost identical, but they start to differ from $z > 0.56$. In the same reference, the authors show that this happens for $z=0.5$.\\
We find that for $z \gg 1$, the parameters $\left\{ q,r,s \right\}  \rightarrow \left \{ 0.88, 2.4, 1.2\right \}$, observing that all three values deviate significantly from the values expected for the standard matter-dominated era. Recalling that in the limiting case ($\Omega_c+\Omega_b= 1$, $\Omega_{de}= 0$), neglecting radiation, the value $s$ predicted by the SCDM model is non-zero ($r=1, s=1$). In this case, $s=1.2$ is far way  from the value $s=1$.
From Fig.(\ref{figur16:2}), we observe that the trajectory In the $(q, r)$ plane is linear and starts in the upper right corner, in the region bounded by $0<q < 1$ and $r>1$.  The curve evolves in the negative direction of $q$ and negative of $r$. At the present, $z =0$, one finds $\left\{ q, r, s \right\} = \left \{ -0.67, 1.25, -0.071\right \}$, and also that $q$ and $r$ are not close values for the Statefinder parameters expected for $\Lambda$CDM, $\left\{ q_0, r_0, s_0 \right\} = \left \{ -0.55, 1, 0\right \}$. 
When $z \rightarrow -1$, the parameters $\left\{ q,r,s \right\}  \rightarrow \left \{-1,1,0 \right \}$. Thus, the parameters present an asymptotic behaviour toward the dS fixed point in the future.\\
In the $(s, r)$ plane, Fig.(\ref{figur16:3}) shows that the curve starts in the upper right corner of the figure, in the region bounded by $s < 0$ and $r > 1$. The trajectory is non-linear in behavior and proceeds in the negative direction of $r$, but in the positive direction of $s$. We corroborate the information provided by the $(q, r)$ plane; the trajectory of the model converges to $\Lambda$CDM. We conclude that this model is incompatible with $\Lambda$CDM, since, although it converges to $\Lambda$CDM, the present values and in the epoch that matter dominates, deviate significantly from $\Lambda$CDM.

\subsubsection{MODEL 17. Asymmetric model. $Q_m=-3H\alpha\rho_m, Q_x=-3H\beta(\rho_m+\rho_x)$}
\begin{figure}[htp]
  \centering
  \label{figur}
   \subfloat[]{\label{figur17:1}\includegraphics[width=80mm]{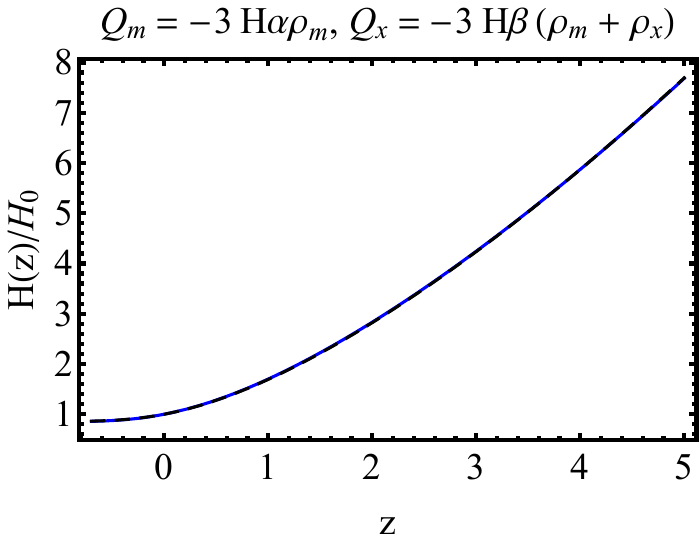}}
   \\
  \subfloat[]{\label{figur17:2}\includegraphics[width=80mm]{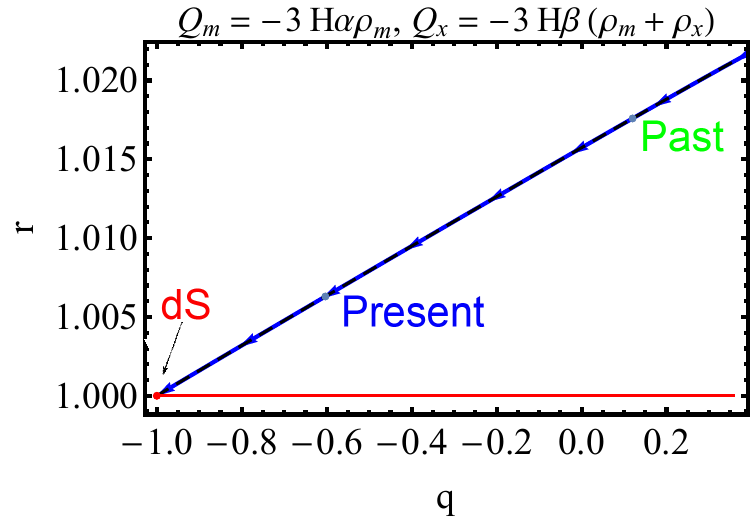}}
  \subfloat[]{\label{figur17:3}\includegraphics[width=80mm]{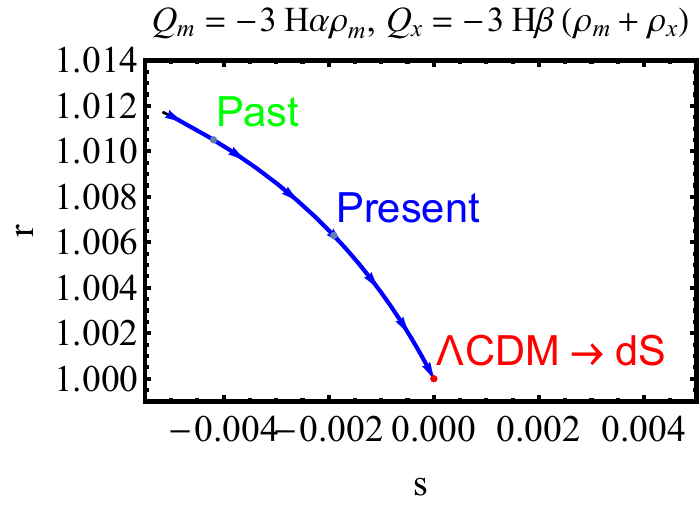}}
\caption{{\small In Figure (a), it is shown the dimensionless Hubble rate $H(z)/H_0$ as a function of redshift $z$ for the Asymmetric model, $Q_m=-3H\alpha\rho_m, Q_x=-3H\beta(\rho_m+\rho_x)$, and we represent the self-interaction (dashed line), as also, the Hubble rate of $\Lambda$CDM model (solid line). For this model, Figures (b) and (c) show the parametric curves in the $q - r$ and $s - r$ planes, respectively.}}  
\end{figure}

MODEL 17 is evaluated numerically using the values: $\Omega_m=0.14 \pm 0.09$, $\Omega_x=0.146 \pm 0.085$, $\alpha =0.09\pm 0.17$ and $\beta =-0.01 \pm0.17$.\\
In Fig. (\ref{figur17:1}), we show the dimensionless Hubble rate, like the previous model, we plot the self-interaction (dashed line), which overlaps with the plot of Hubble rate for the $\Lambda$CDM model (solid line). Unlike the previous model, the $H(z)$ curves of the asymmetric model and the reduced $\Lambda$CDM curve fit better for a larger redshift interval.\\
We find that for $z \gg 1$, the parameters $\left\{ q, r, s \right\}  \rightarrow \left \{0.51, 1.02, 1 \right \}$. It is observed that the values of $q$ and $r$ are very close to the expected values for the standard matter-dominated era. Additionally, the value of $s=1$ corresponds to SCDM ($r=1, s=1$).\\
In the $(q, r)$ plane from Fig.(\ref{figur17:2}), we note that the trajectory is linear and starts in the upper right corner, in the region bounded by $0<q < 1$ and $r<1$.  The curve evolves in the negative direction of $q$ and negative of $r$. At the current time, $z =0$, it is found that $\left\{ q, r, s \right\}  = \left \{-0.6, 1.01, -0.002 \right \}$, and according, $r$ and $s$ are close values for the expected Statefinder parameters for $\Lambda$CDM, but the value of $q(0)$ deviates significantly.
When $z \rightarrow -1$ the parameters $\left\{ q,r,s \right\}  \rightarrow \left \{-1,1,0 \right \}$. Thus, the parameters exhibit asymptotic behaviour towards the dS fixed point in the future, as also does the symmetric model.
In the $(s, r)$ plane, Fig.(\ref{figur17:3}) shows that the curve starts in the upper left corner of the figure, in the region bounded by $s < 0$ and $r > 1$. The trajectory has a non-linear behavior and advances in the negative direction of $r$, and in the positive direction of $s$. We can compare with the information provided by the $(q, r)$ plane, the trajectory of the model converges to $\Lambda$CDM. We conclude that this model is compatible with $\Lambda$CDM. Although the value of $q(0)$ deviates from $\Lambda$CDM, the values $q$, $r$, and $s$ (as a whole) are close to $\Lambda$CDM in the all epochs. In contrast with the Symmetric model, we note that this model is able to reproduce a standard matter-dominated era.
\\
\section{Comparison between models}\label{section6}

\begin{figure}[htp]
  \centering
  \label{figur}
 \subfloat[]{\label{figurcomp:1}\includegraphics[width=90mm]{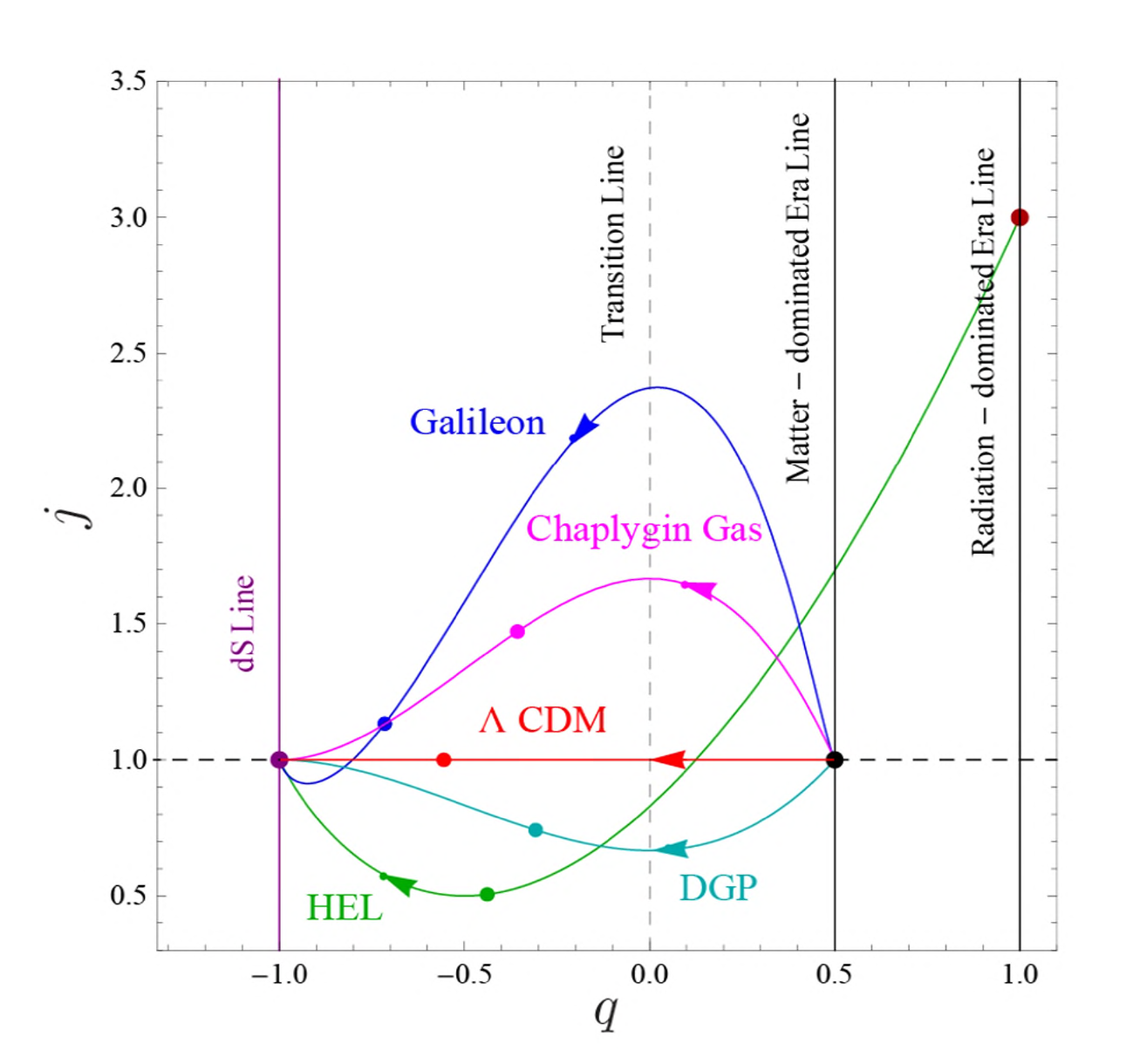}}
 \subfloat[]{\label{figurcomp:2}\includegraphics[width=90mm]{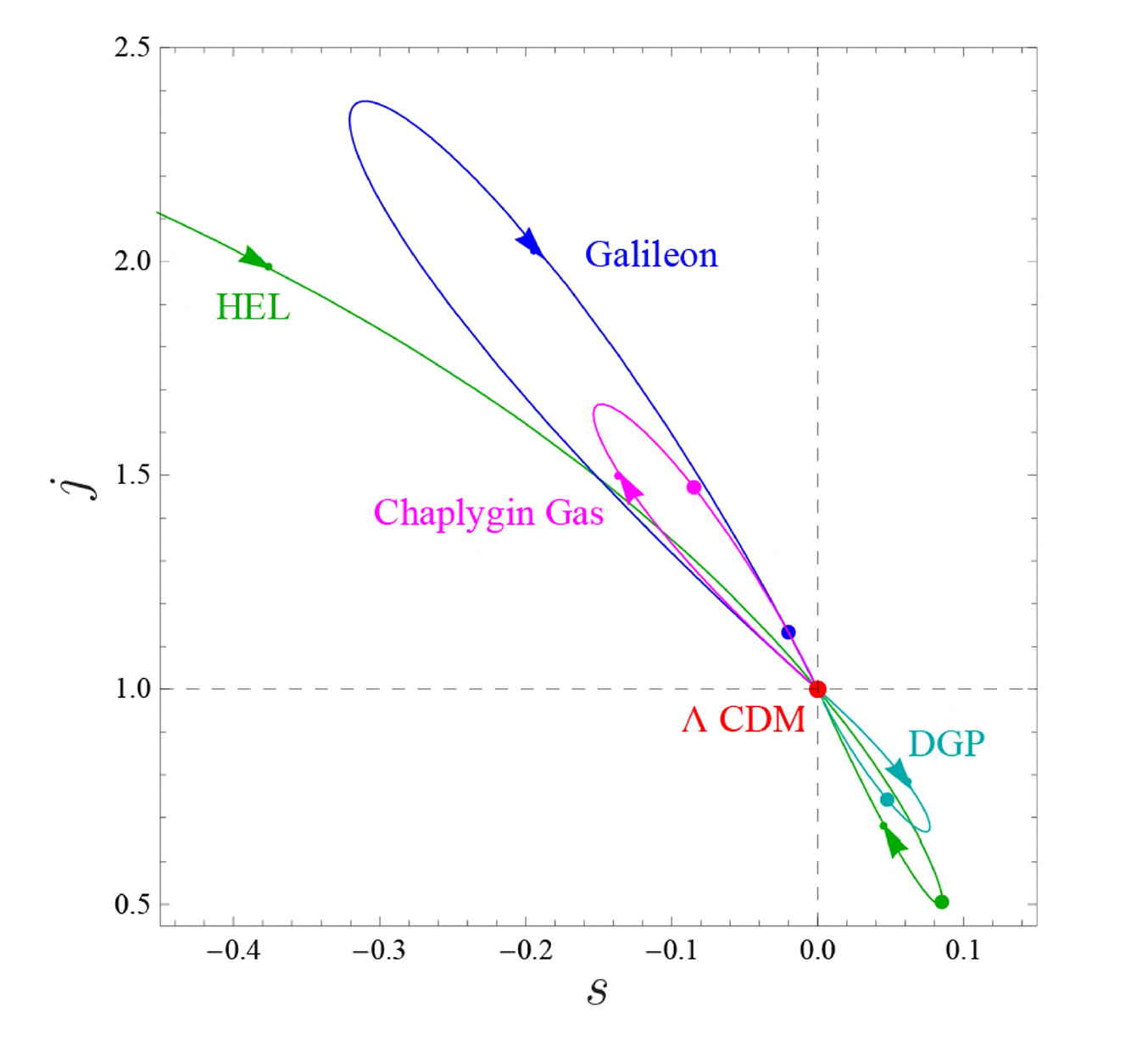}}
\caption{{\small The parametric curves in the $q-j$ plane are shown in Figure (a). The dS state $q = -1$ is shown by the vertical purple line. The plots in the $s-j$ plane are displayed in Figure (b). The HEL model is shown by the green curve in both panels. The $\Lambda$CDM, DGP, Chaplygin gas, and Galileon models are represented by the Red, Cyan, Magenta, and Blue curves, respectively. The $\Lambda$CDM point $(0,1)$ is where horizontal and vertical dashed lines connect. The dots on the curves represent the current values of the  $(s, j)$ or $(q, j)$ pair, while the dark dots on the curves reflect the matter-dominated phases of the models. Figure taken from \cite{Akarsu:2013xha}.}}  
\end{figure}

Now, we are able to compare between the models by looking at the trajectories from where they start. We consider Figs. (\ref{figurcomp:1}) and (\ref{figurcomp:2}) for comparison. In accordance to \cite{Sahni:DMDE} and \cite{Akarsu:2013xha}, in the case of models 7, 16, and 17, it is observed that the pair $(s, r)$ starts on the left-hand side of the fixed point $(0,1)$ corresponding to $\Lambda$CDM, which is characteristic of the hybrid expansion law (HEL), Chaplygin gas and Galileon models, such that $s < 0$ and $r > 1$ \cite{Sami:2012uh,Akarsu:2013xha}. We point out that this behaviour is different from the case of the Quintessence model, for which the trajectory in the $(s, r)$ plane is observed to start in the region $0 < s < 1$ and $r < 1$. \\
On the other hand, we can see that the trajectory of MODEL 7 in Fig. (\ref{figur331:2}), is very similar to the HEL model, which the trajectory in the $(q, r)$ plane starts in the region bounded by $0 < q < 1$ and $r > 1$.\\
The plot for model 10 in the $(s, r)$ plane is very similar to that of models 5 and 9. For the latter model, the deviations are larger than in the similar models 5 and 9.\\
As we established above, for the limiting case ($\Omega_m=1$, $\Omega_{de}=0$, and neglecting radiation), the values of the Statefinder parameters are $r=1, s=1$, corresponding to SCDM. It is observed in the Table (\ref{statevalues}) that when $z \gg 1$, $s$ becomes equal to 1 or very close to 1 obtained for models 1, 3, 5, 7, 8, 13, 15 and 17.
We can conjecture and give an explanation for these values by assuming that it may be since in these models the $Q$ function depends mainly on DM and because the impact of DE is smaller on the interaction, since the analysis takes place in the matter-dominated era.\\
In addition, we can compare the MODEL 7 with a Statefinder analysis performed for running vacuum models (RVM). If we take the plots in $r-s$ plane, we see that Fig.(\ref{figur331:3}) is similar to Fig. 1 of \cite{Panotopoulos:2019xbw}. As noted above, after the present time, the curve $r-s$ generates a twist and closes on itself, converging to the point $(0,1)$. The similarity is that in our MODEL 7 and model I of \cite{Panotopoulos:2019xbw}, since they have a $Q$ rate linearly depending on dark matter density. In contrast, for the DE energy density dependence (model II in \cite{Panotopoulos:2019xbw}), we can see that the curve does not converge to the point $(0,1)$. In model II the asymptotic fixed point is a scaling solution, where the scaling solution is when the energy density of DE depends on an inverse power of the scale factor $a(t)$.

%TABLA DE RESUMEN DE VALORES

%%%%%%%%%%%%%%%%%%%%%%%%%%%%

\begin{table}[]
\centering
\begin{tabular}{|l|l|l|l|l|l|l|l|l|l|l|l|l|}
\hline
$Q$ & $q(z\gg1)$ & $r(z\gg1)$& $s(z\gg1)$& $q(0)$  &$r(0)$  &$s(0)$  &$q(-1)$   &   $r(-1)$       & $s(-1)$                       
\\ 
\hline
$Q_1=3\gamma H \rho_c$  
&0.42&	0.79&	0.93&	-0.50&	0.99&	0.003&	-1.0&	1.1& -0.03\\ 
\hline
$Q_2=3\gamma H \rho_{de}$
&0.5&  	1.&	0.02&	-0.53&	0.94&	0.02&	-0.97&	0.91&	0.02     \\ 
\hline
$Q_3= 3\lambda_{c}H\rho_{c} +3\lambda_{de}H\rho_{de}$
&0.44&	0.84&	0.96&	-0.56&	0.99&	0.005&	-1.0&	1.1& -0.012 \\ 
\hline
$Q_4= 3H \gamma \frac{\rho_{c}\rho_{de}}{\rho}$
&0.50&	1.&	0.06&	-0.52&	0.94&	0.019&	-1.&	1.&	0. \\ 
\hline
$Q_5= 3H \gamma \frac{\rho_{c}^2}{\rho}$
&0.44&	0.84&	0.96&	-0.56&	1.&	-0.019&	-1.&	1.2&	-0.037 \\ 
\hline
$Q_6= 3H \gamma \frac{\rho_{de}^2}{\rho}$
&0.5&	1.&	0.&	-0.51&	0.84&	0.05&	-0.89&	0.69&	0.074 \\ 
\hline
$Q_7= q(\alpha \dot{\rho}_{c} + 3 \beta H\rho_{c})$
&0.51&	1.&	1.0&	-0.59&	0.99&	0.002&	-1.&	1.&	0. \\ 
\hline
$Q_8= q(\alpha \dot{\rho}_{tot} + 3 \beta H\rho_{tot})$  
&0.52&	1.1&	1.0&	-0.59&	0.93&	0.02&	-0.96&	0.89&	0.024 \\ 
\hline
$Q_9= q(\alpha \dot{\rho}_{de} + 3 \beta H\rho_{de})$
&0.5&	1.03&	0.007&	-0.59&	1.02&	-0.008&	-1.02&	1.06&	-0.014 \\ 
\hline
$Q_{10}=3\alpha H (\rho_{de}'+\rho_{c}')$
&0.38&	0.68&	0.86&	-0.73&	2.35&	-0.37&	-1.7&	4.1& -0.47 \\ 
\hline
$f(\tilde{r})=1$,
$Q= 3H \gamma \frac{\rho_{c}\rho_{de}}{\rho_{c}+\rho_{de}}$
&0.5&	1.&	0.06&	-0.52&	0.94&	0.019&	-1.&	1.&	0. \\ 
\hline
$f(\tilde{r})=\frac{1}{\tilde{r}}$,
$Q= 3H \gamma \frac{\rho_{de}^2}{\rho_{c}+\rho_{de}}$
&0.5&	1.&	0&	-0.51&	0.84&	0.05&	-0.89&	0.69&	0.074\\ \hline
$f(\tilde{r})=\tilde{r}$,
$Q= 3H \gamma \frac{\rho_{c}^2}{\rho_{c}+\rho_{de}}$      
&0.44&	0.84&	0.96&	-0.52&	0.98&	0.006&	-1.&	1.&	0. \\ \hline
$f(\tilde{r})=1+\frac{1}{\tilde{r}}$,
$Q= 3H \gamma \rho_{de}$                                                                                         
&0.5&	1&	0.037&	-0.52&	0.89&	0.037&	-0.94&	0.84&	0.037 \\ \hline
$f(\tilde{r})=1+\tilde{r}$,
$Q= 3H \gamma \rho_{c}$                                                                                
&0.5&	0.991&	1&	-0.54&	0.99&	0.001&	-1.&	1.&	0. \\ 
\hline
\begin{tabular}[c]{@{}l@{}}SELF-INTERACTION\\ Symmetric model\\ 
$Q_m=-3H\alpha\rho_m$,
\\ 
$Q_x=-3H\beta\rho_x$
\end{tabular}   
&0.88&	2.4&	1.2&	-0.67&	1.25&	-0.071&	-1.&	1.&	0.  \\ \hline
\begin{tabular}[c]{@{}l@{}}SELF-INTERACTION\\ Asymmetric model\\ 
$Q_m=-3H\alpha\rho_m$,
\\ 
$Q_x=-3H\beta(\rho_m+\rho_x)$
\end{tabular} 
&0.51&	1.02&	1.&	-0.6&	1.006&	-0.002&	-1.&	1.&	0. \\ 
\hline
\end{tabular}
\caption{Summary of present and asymptotic values of Statefinder parameters for all the interacting models studied in the present research.}
\label{statevalues}
\end{table}

%%%%%%%%%%%%%%%%%%%%%%%%%

\section{Conclusions}\label{section7}

The nature of the components of the dark sector of our Universe is still unknown. Several models have been proposed in order to explain the accelerating expansion of the Universe. In this context, interacting DE models have been proposed to solve or alleviate 
the emerging cosmological tensions of $\Lambda$CDM model as well as the cosmic coincidence problem.\\
In this paper, we have used the Statefinder diagnostic with the purpose to discriminate between several interacting DE models. In particular. we have investigated 17 interacting models between DE and DM, which have been already studied in the recent literature and constrained with astronomical and cosmological data. Specifically, the investigated models belong to the following categories:
following categories:
i) linear models in energy densities of DE and DM, 
ii) non-linear models, 
iii) models with a change of direction of energy transfer between DE and DM, 
iv) models involving derivatives of the energy densities, 
v) parametrized interactions through a function of the coincidence parameter $\tilde{r}$, and finally we also consider two kinds of models with a self-interaction between DM, without DE. \\
The models we have chosen allow us to solve the conservation differential equations and to find an analytical expression for the dimensionless Hubble rate $E(z)=H(z)/H_0$. Thus, by using $E(z)$ we have computed the expressions for the Statefinder parameters $q(z)$, $r(z)$, and $s(z)$. Then, these parameters were evaluated for the matter-dominated era ($z \gg 1$), at the present epoch ($z=0$), and in the future ($z \rightarrow -1$). These values are shown in Table (\ref{statevalues}), as a summary. It was necessary to plot in the $q-r$ and $s-r$ planes the respective parametric curves for each case. The trajectory of each model was compared with the points associated to $\Lambda$CDM, i.e., $(0,1)$ in the $s-r$ plane and the dS fixed point $(-1,1)$ in the $q-r$ plane. Additionally, we found that the "anomalous" behavior of the parameter $s$ for $z \gg 1$, corresponds to the point $(1,1)$, associated to the SCDM model.\\
Differences and similarities were found between models of the same category (linear, non-linear, etc.), depending on the deviations obtained with respect to $\Lambda$CDM. Comparing all of these models, some of them converge to $\Lambda$CDM at late times, such as models 4, 7, 11, 13, 15 and the self-interacting models 16 and 17. In that sense, the worst performing model in the future is MODEL 10, since its deviations are very large with respect to the others.\\
In the current epoch, the models 5, 3 and 15 show which are values very close to those predicted by $\Lambda$CDM, i.e., $\{q_0, r_0, s_0\} = \{-0.55,1,0\}$. But, to a lesser extent, models 13, 7, 9 and 17 are close to $\Lambda$CDM. On the other hand, the worst behavior corresponds to MODEL 10, and the symmetric model, since these deviate significantly for the current epoch, especially the parameter $q$.\\
In models 2, 4, 6, 9, 11, 12 and 14 the values found for the three Statefinder parameters behave very similar to $\Lambda$CDM in the matter-dominated era, i.e., $\{q(z \gg 1), r( z \gg 1), s( z \gg 1)\} = \{0.5,1,0\}$. At this same epoch, models 1, 3, 5, 7, 8, 13, 15, and the asymmetric model predict values of $s$ close to 1, which we established is related to SCDM. We inferred that this was due to the larger impact of DM in the interactions, given the form of the $Q$ rate and the small contribution of the DE energy density. Furthermore, at this epoch, we see that models 16, 10, 1, 3, 5, and 13 deviate significantly from the $q$ and $r$ values predicted by $\Lambda$CDM.\\
The models parameterized and their equivalents have similar behaviours, except 5-13 and 1-15. Actually, we found that models 5 and 1 do not converge to $\Lambda$CDM in the future, but 13 and 15 do instead. As we said, this may be due to the fact that the parameterized models use more updated data and $w=-1$. This leads us to conclude that the models analyzed by Statefinder analysis cannot be definitively discarded, as more updated data can lead to better model performance. \\
As we have shown, we can compare our models with other different DE models extensively studied in the literature, such as HEL, Chaplygin gas, Galileon models, Quintessence , and DGP from \cite{Akarsu:2013xha}. This can be done by considering the deviations from $\Lambda$CDM and the shape of the trajectories in the $(s, r)$ plane, as we look at the origins of the curves. \\
For future research, we would like to complement this work and to corroborate the convergence of the models by analyzing dynamical systems and observing the stability of the fixed points, as has been done for other DE models \cite{Bahamonde:2017ize,Panotopoulos:2019xbw}.\\
This work opens up the possibility to further investigation on others interacting DE models that have a more complex expression for $Q$, for example, where the EoS parameter $w$ or strengths of the couplings depend on redshift or time.  We hope to be able to address this
point in a future work.

%Thus, if it is possible to find the solutions for the densities and hence analytical expression for $E(z)$, then we think Statefinder analysis can be easily applied. 

%Even tentative and challenging work could attempt to apply Statefinder analysis to models where $H(z)$ is found numerically. For instance, MODEL 7, in which $Q$ is proportional to $q$, it exhibits good behavior in all three epochs, seems to be compatible with $\Lambda$CDM. The predictions of this model could be improved by considering the time-derivative terms in the expression for $Q$. However, the transcendental differential equation for $H(z)$ would most likely have to be solved numerically.

\section{Acknowledgments}

A.R. acknowledge financial support from the Generalitat Valenciana through PROMETEO PROJECT CIPROM/2022/13.  A.R. is funded by the Maria Zambrano contract ZAMBRANO 21-25 (Spain). 
N. Videla and J. Saavedra acknowledge the financial support of Fondecyt Grant 1220065.

%
% For  figures use
%\begin{figure*}
%\vspace*{5cm}       % Give the correct figure height in cm
%\includegraphics{leer.eps}
%\end{figure*}
% or  this
%\begin{figure}
%\centering
%\resizebox{0.75\textwidth}{!}{%
%  \includegraphics{leer.eps}
%\vspace{5cm}       % Give the correct figure height in cm
%\end{figure}
%\begin{table}
%\centering
%\begin{tabular}{lll}
%\hline\noalign{\smallskip}
%\end{tabular}
% Or use
%\vspace*{5cm}  % with the correct table height
%\end{table}

%
% BibTeX users please use
% \bibliographystyle{}
 \bibliography{References}
% \begin{thebibliography}{}
% \bibliography{References}
% \end{thebibliography}

\end{document}